\documentclass[preprint,showpacs,preprintnumbers,nofootinbib,amsmath,amssymb]{revtex4}
\usepackage{graphicx}
\usepackage{dcolumn}
\usepackage{bm}
\begin{document}
\title{Pairing within the self-consistent quasiparticle
random-phase approximation at finite temperature}
\author{N. Dinh Dang$^{1, 2}$}
 \email{dang@riken.jp}
 \author{N. Quang Hung$^{1}$}
 \altaffiliation[On leave of absence from the ]{Institute of Physics and Electronics, Hanoi, Vietnam}
  \email{nqhung@riken.jp}
\affiliation{1) Heavy-Ion Nuclear Physics Laboratory, RIKEN Nishina Center
for Accelerator-Based Science,
2-1 Hirosawa, Wako City, 351-0198 Saitama, Japan\\
2) Institute for Nuclear Science and Technique, Hanoi, Vietnam}
\date{\today}
\begin{abstract}
An approach to
pairing in finite nuclei at nonzero
temperature is proposed, which incorporates
the effects due to the quasiparticle-number fluctuation (QNF) around
Bardeen-Cooper-Schrieffer (BCS)
mean field and dynamic coupling to quasiparticle-pair vibrations
within the self-consistent quasiparticle random-phase approximation
(SCQRPA).
The numerical calculations of pairing gap, total
energy, and heat capacity were carried out within a
doubly folded multilevel model as well as
realistic nuclei $^{56}$Fe and $^{120}$Sn.
The results obtained show that, under the effect of QNF,
in the region of moderate and strong couplings, the sharp
transition between the superconducting and normal phases is
smoothed out, resulting in a thermal pairing gap, which
does not collapse at the BCS critical temperature, but has a
tail, which extends to high temperature. The dynamic coupling of
quasiparticles to SCQRPA vibrations significantly improves the
agreement with the results of exact calculations and those obtained
within the finite-temperature quantal Monte Carlo method for the total
energy and heat capacity.
It also causes a deviation of the quasiparticle occupation numbers
from the Fermi-Dirac distributions for free fermions.
\end{abstract}

\maketitle
\section{INTRODUCTION}
\label{Intro}
Pairing phenomenon is a common feature in
strongly interacting many-body systems ranging from tiny ones such as
atomic nuclei to very large ones such as neutron stars.
Because of its simplicity,
the Bardeen-Cooper-Schrieffer (BCS) theory~\cite{bcs},
which explains the conventional superconductivity,
has been widely employed as the first step
in nuclear structure calculations
that include pairing forces. In infinite systems such as
low-temperature superconductors,
the BCS theory offers a correct description of the pairing
gap as functions of temperature $T$ and pairing-interaction strength $G$.
Here, as $T$ increases, the BCS gap decreases from its value $\Delta(0)$ at $T=$ 0
until it collapses at a
critical temperature $T_{\rm c}=$ 0.567$\Delta(0)$, at which
the phase transition between the superconducting phase and normal
one (SN-phase transition) occurs~\cite{es,Landau}.
However, the application of the BCS theory to small systems such as
atomic nuclei needs to be carried out with a certain care
since quantal and
thermal fluctuations are not negligible in finite systems,
especially when the number of particles is small.

The effects of thermal fluctuations on the pairing
properties of nuclei have been the subject of numerous theoretical studies
in the last three decades. In the seventies, by applying the macroscopic
Landau theory of phase transitions to a uniform model, Moretto has shown that
thermal fluctuations smooth out the sharp SN phase transition in
finite systems~\cite{Moretto}. In the eighties, this approach was incorporated by
Goodman into
the Hartree-Fock-Bogoliubov (HFB) theory at finite
temperature~\cite{Goodman1} to
account for the effect of thermal fluctuations~\cite{Goodman2}. Theoretical studies within the static-path approximation
(SPA) carried out in the nineties also came to the
non-vanishing pairing correlations at finite temperature~\cite{SPA}, which are
qualitatively similar to the predictions by Landau theory of phase
transitions. The shell-model and Monte-Carlo shell-model
calculations~\cite{shell,Monte} also show that pairing does not
abruptly vanish
at $T_{\rm c}$, but still survives at $T>T_{\rm c}$. For rotating
systems, Frauendorf and collaborators have
recently shown a phenomenon of pairing induced by
temperature~\cite{Frau}, which reflects strong fluctuations of the
order parameter in very small systems with a fixed number of particles.
The recent microscopic approach to thermal pairing, called modified-HFB (MHFB)
theory~\cite{MHFB},
includes the quasiparticle-number fluctuation (QNF) in
the modified single-particle density
matrix and particle-pairing tensor.
Its limit of constant pairing interaction $G$
is the modified BCS (MBCS) theory~\cite{MBCS1,MBCS2,MBCS3,MBCS4}. The
MBCS theory predicts a pairing gap, which does not collapse at
$T_{\rm c}$, but monotonously decreases with increasing $T$, in qualitative
agreement with the predictions by the Landau theory of phase
transitions and SPA. This feature also agrees with the results
obtained by averaging the exact eigenvalues of the pairing problem
over the canonical ensemble with a temperature-dependent partition
function~\cite{MBCS3}. The recent extraction of pairing gap from the
experimental level densities~\cite{exp} confirms that the pairing
gap does not vanish at $T_{\rm c}$ but decreases as $T$ increases, in
line with the predictions by these approaches.

The above mentioned approaches are based on the
independent quasiparticles, whose occupation numbers follow
the Fermi-Dirac distribution of free fermions. Dynamic effects such as those due to
coupling to small-amplitude vibrations within the random-phase
approximation (RPA) are ignored. These effects have
recently been explored by extending the self-consistent
particle-particle RPA (SCRPA) to finite temperature using the double-time
Green's function method~\cite{DaTa}.
Since the SCRPA fails in the
region of strong pairing, where it should be replaced by the
quasiparticle representation, it is highly desirable to develop
a self-consistent quasiparticle RPA
(SCQRPA) at finite temperature, which is workable with any value of
pairing interaction parameter $G$.

Recently, we have  developed in Ref. \cite{SCQRPA}
a SCQRPA for the multilevel pairing
Hamiltonian and applied it to the Richardson model~\cite{Ric} at
zero temperature. The derivation of the SCQRPA is based
on a set of renormalized BCS equations, which include the corrections
due to the QNF and the
SCQRPA. The latter arise from the expectation values $\langle{\cal A}^{\dagger}_{j}{\cal
A}^{\dagger}_{j'}\rangle$ and $\langle{\cal A}^{\dagger}_{j}{\cal
A}_{j'}\rangle$ in the correlated ground state. Here ${\cal
A}^{\dagger}_{j}$ is
the product of two time-reversal conjugated quasiparticle
operators, $\alpha^{\dagger}_{j}$ and $\alpha^{\dagger}_{-j}$,
corresponding to the $j$-th orbital. Within the particle-particle
($pp$) SCRPA~\cite{SCRPA}, these expectation values overscreen the attractive
pairing interaction, turning it into repulsion in agreement with the
trend of the exact solutions of the Richardson model. For this
reason, the expectation values $\langle{\cal A}^{\dagger}_{j}{\cal
A}^{\dagger}_{j'}\rangle$ and $\langle{\cal A}^{\dagger}_{j}{\cal
A}_{j'}\rangle$ are called the screening factors.
The goal of the present study is to
extend the SCQRPA in Ref. \cite{SCQRPA} to non-zero temperature
to explore the effects due to QNF as well as coupling
to QRPA vibrations on the pairing properties of finite systems in a
self-consistent way.

The article is organized as follows. The derivation of the
equations for quasiparticle propagation, which include the
effects of QNF and SCQRPA corrections
as well as coupling of quasiparticles to
pair vibrations at finite temperature is
presented Section \ref{formalism}. Two approximation schemes
will be considered, which
are based on the thermal quasiparticle representation without and
including dynamic coupling to SCQRPA quasiparticle-pair vibrations.
In Section \ref{results},
the developed approach undergoes a thorough numerical test
within the Richardson model as well as in realistic nuclei $^{56}$Fe
and $^{120}$Sn.
The last section summarizes the article, where conclusions are drawn.
\section{FORMALISM}
\label{formalism}
\subsection{Quasiparticle Hamiltonian}
\label{Hamiltonian}
The pairing Hamiltonian
    \begin{equation}
    H=\sum_{jm}\epsilon_{j}a_{jm}^{\dagger}a_{jm}-
    G\sum_{jj'}\sum_{mm'>0}a_{jm}^{\dagger}a_{j\widetilde{m}}^{\dagger}
    a_{j'\widetilde{m'}}a_{j'm'}~.
    \label{H}
    \end{equation}
    describes a set of $N$ particles with single-particle energies
    $\epsilon_{j}$, which are generated by particle creation operators
    $a_{jm}^{\dagger}$ on $j$-th orbitals with shell degeneracies
    $2\Omega_{j}$ ($\Omega_{j}=j+1/2$), and interacting via a
    monopole-pairing force with a constant parameter $G$.
    The symbol $~~\widetilde{}~~$ denotes
    the time-reversal operator, namely
    $a_{j\widetilde{m}}=(-)^{j-m}a_{j-m}$.
    In general, for a two-component system with $Z$ protons and $N$
    neutrons, the sums in Eq. (\ref{H}) run over all $j_{\tau}m_{\tau}$,
    $j'_{\tau}m'_{\tau}$, and $G_{\tau}$ with $\tau=(Z,N)$. This
    general notation is omitted here as the calculations in the
    present article are carried out only for one type of particles.

    By using the Bogoliubov's transformation from the particle operators,
    $a_{jm}^{\dagger}$ and $a_{jm}$, to the
    quasiparticle ones, $\alpha_{jm}^{\dagger}$ and $\alpha_{jm}$,
    \begin{equation}
    a_{jm}^{\dagger}=u_j\alpha_{jm}^{\dagger} +
    v_j\alpha_{j\widetilde{m}}~, \hspace{5mm}
    a_{j\widetilde{m}}=u_j\alpha_{j\widetilde{m}}
    -v_j\alpha_{jm}^{\dagger}~,
    \label{Bogo}
    \end{equation}
    the pairing Hamiltonian (\ref{H}) is transformed into the
    quasiparticle Hamiltonian as follows~\cite{MBCS2,MBCS3}
    \[
    H=a+\sum_j{b_j\mathcal{N}_j}+\sum_j{c_j(\mathcal{A}_j^{\dagger}+\mathcal{A}_j})
    +\sum_{jj'}{d_{jj'}\mathcal{A}_j^{\dagger}\mathcal{A}_{j'}}
    +\sum_{jj'}{g_j(j')(\mathcal{A}_{j'}^{\dagger}\mathcal{N}_j+\mathcal{N}_j\mathcal{A}_{j'})}
    \]
    \begin{equation}
    +\sum_{jj'}{h_{jj'}(\mathcal{A}_j^{\dagger}\mathcal{A}_{j'}^{\dagger}
    +\mathcal{A}_{j'}\mathcal{A}_j)}+\sum_{jj'}{q_{jj'}\mathcal{N}_j\mathcal{N}_{j'}}~,
    \label{Hqp}
    \end{equation}
    where $\mathcal{N}_{j}$ is the quasiparticle-number operator,
    whereas
    $\mathcal{A}_j^{\dagger}$ and $\mathcal{A}_j$ are the creation and
    destruction operators of a pair of time-reversal conjugated quasiparticles:
    \begin{equation}
    \mathcal{N}_j=\sum_{m=-\Omega_{j}}^{\Omega_{j}}
    \alpha_{jm}^{\dagger}\alpha_{jm} = \sum_{m=1}^{\Omega_{j}}(\alpha_{jm}^{\dagger}\alpha_{jm}
    +\alpha_{j-m}^{\dagger}\alpha_{j-m})
    \label{Ncal}~,
    \end{equation}
    \begin{equation}
    \mathcal{A}_j^{\dagger} = \frac{1}{\sqrt{2}}\big[\alpha^{\dagger}_{j}
    \otimes\alpha_{j}^{\dagger}\big]_{0}^{0} = \frac{1}{\sqrt{\Omega_{j}}}\sum_{m=1}^{\Omega_{j}}
    \alpha_{jm}^{\dagger}\alpha_{j\widetilde{m}}^{\dagger}~,
    \hspace{5mm}
    \mathcal{A}_j=(\mathcal{A}_j^{\dagger})^{\dagger}~.\label{A}
    \end{equation}
    They obey the following commutation relations
    \begin{eqnarray}
    &&[\mathcal{A}_j~,~\mathcal{A}_{j'}^{\dagger}] = \delta_{jj'}{\cal D}_{j}~,\hspace{2mm}{\rm where} \hspace{2mm}
    {\cal D}_{j}=1-\frac{\mathcal{N}_j}{\Omega_{j}}~,
    \label{[AA]}\\
    &&[\mathcal{N}_j~,~\mathcal{A}_{j'}^{\dagger}] = 2\delta_{jj'}\mathcal{A}_{j'}^{\dagger}~,
    \hspace{5mm}
    [\mathcal{N}_j~,~\mathcal{A}_{j'}] = -2\delta_{jj'}\mathcal{A}_{j'}~.
    \label{[NA]}
    \end{eqnarray}
    The functionals
    $a$, $b_j$, $c_j$, $d_{jj'}$, $g_j(j')$, $h_{jj'}$, $q_{jj'}$
    in Eq. (\ref{Hqp}) are given in terms of the coefficients $u_j$, $v_j$
    of the
    Bogoliubov's transformation, and the single particle
    energies $\epsilon_j$ as (See Eqs. (7) -- (13) of Ref.
    \cite{MBCS2}, e.g.)
    \begin{equation}
    a=2\sum_{j}\Omega_{j}\epsilon_{j}v_{j}^{2}
    -G\big(\sum_{j}\Omega_{j}u_{j}v_{j}\big)^{2}-G\sum_{j}\Omega_{j}v_{j}^{4}~,
    \label{a}
    \end{equation}
    \begin{equation}
    b_{j}=\epsilon_{j}(u_{j}^{2}-v_{j}^{2})+2Gu_{j}v_{j}\sum_{j'}\Omega_{j'}u_{j'}v_{j'}
    +Gv_{j}^{4}~,
    \label{b}
    \end{equation}
    \begin{equation}
    c_{j}=2\sqrt{\Omega_{j}}\epsilon_{j}u_{j}v_{j}-G\sqrt{\Omega_{j}}
    (u_{j}^{2}-v_{j}^{2})\sum_{j'}\Omega_{j'}u_{j'}v_{j'}-
    2G\sqrt{\Omega_{j}}u_{j}v_{j}^{3}~,
    \label{c}
    \end{equation}
    \begin{equation}
    d_{jj'}=-G\sqrt{\Omega_{j}\Omega_{j'}}(u_{j}^{2}u_{j'}^{2}+v_{j}^{2}v_{j'}^{2})=d_{j'j}~,
    \label{d}
    \end{equation}
    \begin{equation}
    g_{j}(j')=Gu_{j}v_{j}\sqrt{\Omega_{j'}}(u_{j'}^{2}-v_{j'}^{2})~,
    \label{g}
    \end{equation}
    \begin{equation}
    h_{jj'}=\frac{G}{2}\sqrt{\Omega_{j}\Omega_{j'}}
    (u_{j}^{2}v_{j'}^{2}+v_{j}^{2}u_{j'}^{2})=h_{j'j}~,
    \label{h}
    \end{equation}
    \begin{equation}
    q_{jj'}=-Gu_{j}v_{j}u_{j'}v_{j'}=q_{j'j}~.
    \label{q}
    \end{equation}
    By setting $\Omega_{j}=$ 1 in Eqs. (\ref{a}) -- (\ref{q}), one
    recovers the expressions for the case with $\Omega$
    doubly-folded levels of the Richardson model (See, e.g., Eqs. (12) -- (18)
    of Ref. \cite{SCQRPA}).
\subsection{Gap and number equations}
\label{Tzer0}
The derivation of the equation for the pairing gap that include the effect
of correlations in the ground state has been presented briefly in Ref.
\cite{SCQRPA} for the Richardson model. For the clarity of the
extension to finite temperature $T$, we
give below the detailed derivation of the gap equation,
which is applied to the more general
quasiparticle Hamiltonian (\ref{Hqp}) and valid for $T\neq$ 0.

The coefficients $u_{j}$ and $v_{j}$ of the Bogoliubov's transformation
(\ref{Bogo}) are determined by using the variational procedure, which
minimizes the expectation value
of the
Hamiltonian ${\cal H}=H-\lambda\hat{N}$ in the
grand canonical ensemble. This
leads to the variational equations~\cite{Schuck}
\begin{equation}
    \frac{\partial{\langle{\cal H}\rangle}}{\partial{u_{j}}}
    +\frac{\partial{\langle{\cal H}\rangle}}{\partial{v_{j}}}
    \frac{\partial{v_{j}}}{\partial{u_{j}}}\equiv\langle [{\cal H},{\cal
    A}_{j}^{\dagger}]\rangle=0~,
        \label{var}
    \end{equation}
    where $\langle\hat{\cal O}\rangle$ denotes the ensemble average of
    the operator $\hat{\cal O}$,
    \begin{equation}
        \langle\hat{\cal O}\rangle\equiv\frac{{\rm Tr}
        [\hat{\cal O}e^{-\beta {\cal H}}]}{{\rm Tr}e^{-\beta{\cal
        H}}}~,\hspace{10mm} \beta=T^{-1}~.
        \label{Trace}
        \end{equation}
    The commutation relation $[{\cal H},{\cal A}_{j}^{\dagger}]$ is
    found by using Eqs. (\ref{[AA]}) and (\ref{[NA]}) as
    \[
        [{\cal H},{\cal A}_{j}^{\dagger}] =
        2b_{j}'{\cal A}_{j}^{\dagger}+ \bigg\{c_{j}'+\sum_{j'}\big[d_{jj'}{\cal A}_{j'}^{\dagger}+
        g_{j'}(j){\cal N}_{j'}+h_{jj'}{\cal A}_{j'}\big]\bigg\}{\cal D}_{j}
        \]
\begin{equation}
    +2\sum_{j'}\bigg\{g_{j}(j')\big[{\cal A}^{\dagger}_{j'}{\cal A}^{\dagger}_{j}+
    {\cal A}^{\dagger}_{j}{\cal A}_{j'}\big]
    +q_{jj'}\big[{\cal A}^{\dagger}_{j}{\cal N}_{j'}
    +{\cal N}_{j'}{\cal A}^{\dagger}_{j}\big]\bigg\}
    +\sum_{j'}h_{jj'}{\cal D}_{j}{\cal
    A}_{j'}~.
        \label{[HA]}
        \end{equation}
The ensemble average of the
commutation relation (\ref{[HA]}) is then
given as
\begin{equation}
    \langle[{\cal H},{\cal
        A}_{j}^{\dagger}]\rangle
=c_{j}'\langle {\cal D}_{j}\rangle
+\sum_{j'}\bigg\{2g_{j}(j')\big[\langle
{\cal A}^{\dagger}_{j'}{\cal A}^{\dagger}_{j}\rangle+
    \langle{\cal A}^{\dagger}_{j}{\cal A}_{j'}\rangle\big]
    +g_{j'}(j)\langle{\cal N}_{j'}{\cal D}_{j}\rangle\bigg\}~,
    \label{<[HA]>}
    \end{equation}
    where the functionals $b_{j}'$ and $c_{j}'$ are
    \begin{equation}
        b_{j}'=b_{j}-\lambda(u_{j}^{2}-v_{j}^{2})~,\hspace{5mm}
        c_{j}'=c_{j}-2\lambda\sqrt{\Omega_{j}} u_{j}v_{j}~,
        \label{b'c'}
        \end{equation}
        i.e. they have the same form as that of
       $b_{j}$ in Eq. (\ref{b}), and $c_{j}$ in Eq.
       (\ref{c}), but with $\epsilon_{j}-\lambda$ replacing
       $\epsilon_{j}$ at the right-hand sides.
    Inserting the explicit expressions for the functionals $c_{j}'$
    from Eq. (\ref{b'c'}) as well as
    $g_{j}(j')$ and $g_{j'}(j)$ from Eq. (\ref{g})
    into the right-hand side of Eq. (\ref{<[HA]>}),
    and equalizing the obtained result to zero as
    required by the variational procedure (\ref{var}), we come to the
    following equation, which is formally identical to the BCS one:
    \begin{equation}
        2(\epsilon'_{j}-Gv_{j}^{2}-\lambda)u_{j}v_{j}
        -\Delta_{j}(u_{j}^{2}-v_{j}^{2})=0~,
        \label{SCBCS}
        \end{equation}
    where, however, the single-particle energies $\epsilon'_{j}$ are renormalized as
    \begin{equation}
        \epsilon_{j}'=\epsilon_{j}+\frac{G}{\sqrt{\Omega_{j}}\langle{\cal D}_{j}\rangle}
       \sum_{j'}\sqrt{\Omega_{j'}}(u_{j'}^{2}-v_{j'}^{2})\bigg
       (\langle{\cal A}_{j}^{\dagger}{\cal A}_{j'}^{\dagger}\rangle
       +\langle{\cal A}_{j}^{\dagger}{\cal A}_{j'}\rangle\bigg)~.
       \label{rene}
       \end{equation}
      The pairing gap
       is found as the solution of the following equation
    \begin{equation}
       \Delta_{j}=\frac{G}{\langle{\cal D}_{j}\rangle}
       {\sum_{j'}\Omega_{j'}\langle{\cal D}_{j}{\cal D}_{j'}\rangle}
       u_{j'}v_{j'}~,
       \label{scgap}
       \end{equation}
       which is level-dependent.
       The coefficients $u_{j}$ and $v_{j}$ of the Bogoliubov's transformation
(\ref{Bogo}) are derived in a standard way from Eq. (\ref{SCBCS}) and the unitarity
constraint $u_{j}^{2}+v_{j}^{2}=$ 1. They read
\begin{equation}
    u_{j}^{2}=\frac{1}{2}\bigg(1
    +\frac{\epsilon'_{j}-Gv_{j}^{2}-\lambda}{E_{j}}\bigg)~,
    \hspace{5mm}
    v_{j}^{2}=\frac{1}{2}\bigg(1
    -\frac{\epsilon'_{j}-Gv_{j}^{2}-\lambda}{E_{j}}\bigg)~,
    \label{uv}
    \end{equation}
    where $E_{j}$ are the quasiparticle energies
    \begin{equation}
        E_{j}=\sqrt{(\epsilon'_{j}-Gv_{j}^{2}
        -\lambda)^{2}+\Delta_{j}^{2}}~.
        \label{Ej}
        \end{equation}
        The particle-number equation
        is obtained by transforming the
       particle-number operator
       $\hat{N}\equiv\sum_{jm}a_{jm}^{\dagger}a_{jm}$ into the
       quasiparticle presentation using the Bogoliubov's
       transformation (\ref{Bogo}) and taking the ensemble
       average. The result is
        \begin{equation}
       N=2\sum_{j}\Omega_{j}\bigg[v_{j}^{2}\langle{\cal
       D}_{j}\rangle +\frac{1}{2}\big(1-\langle{\cal
       D}_{j}\rangle\big)\bigg]~.
       \label{N}
       \end{equation}
The pairing gap $\Delta_{j}$ and chemical potential $\lambda$, which is the Lagrangian
        multiplier in the variational equations (\ref{var}), are
        determined as solutions of Eqs. (\ref{scgap}) and (\ref{N}).

       The right-hand side of Eq. (\ref{scgap}) contains the
       expectation values $\langle{\cal D}_{j}
{\cal D}_{j'}\rangle$, whose exact treatment is not possible
as it involves an infinite boson expansion series~\cite{Samba}.
In the present article, following the treatment on Ref. \cite{SCQRPA},
we use the exact
relation
    \begin{equation}
        \langle{\cal D}_{j}{\cal D}_{j'}\rangle=\langle{\cal D}_{j}\rangle
        \langle{\cal D}_{j'}\rangle + \frac{\delta{\cal N}_{jj'}}
        {\Omega_{j}\Omega_{j'}}~,
        \hspace{5mm} {\rm with}\hspace{5mm}\delta{\cal N}_{jj'} = \langle{\cal N}_{j}{\cal
        N}_{j'}\rangle - \langle{\cal N}_{j}\rangle\langle{\cal
        N}_{j'}\rangle~,
        \label{DD}
        \end{equation}
and the mean-field contraction for the term $\delta{\cal
N}_{jj'}$
\begin{equation}
    \delta{\cal N}_{jj'}\simeq
    2\Omega_{j}\delta{\cal N}_{j}^{2}\delta_{jj'}~,\hspace{5mm}
    \delta{\cal N}_{j}^{2}\equiv n_{j}(1-n_{j})~,
    \label{QNF}
        \end{equation}
        with the quasiparticle occupation number $n_{j}$
        \begin{equation}
       n_{j}=\frac{\langle{\cal
       N}_{j}\rangle}{2\Omega_{j}}=\frac{1}{2}(1-\langle{\cal
       D}_{j}\rangle)~,
       \label{nj}
       \end{equation}
to rewrite the gap equation (\ref{scgap}) as a sum of a level-independent part,
$\Delta$, and a level-dependent part, $\delta\Delta_{j}$, namely
\begin{equation}
    \Delta_{j}=\Delta + \delta\Delta_{j}~,
    \label{gap1}
    \end{equation}
    where
    \begin{equation}
        \Delta = G\sum_{j'}\Omega_{j'}\langle{\cal D}_{j'}\rangle
        u_{j'}v_{j'}~,\hspace{5mm}
        \delta\Delta_{j} = 2G\frac{\delta{\cal N}_{j}^{2}}{\langle{\cal
        D}_{j}\rangle}u_{j}v_{j}~.
\label{gap2}
\end{equation}
The quantity $\delta{\cal N}_{j}^{2}$ in Eqs. (\ref{QNF}) and (\ref{gap2}),
is nothing but the standard expression for
the QNF corresponding to the $j$-th orbital~\cite{Goodman2,MHFB}
\footnote{The
definition (\ref{QNF}) for the QNF $\delta{\cal N}_{j}^{2}$ is
different from that in Eq. (32) of Ref. \cite{SCQRPA} by a factor 2 as
this factor is now put in front of $\Omega_{j}$ to
have the complete shell degeneracy $2\Omega_{j}$.}.
 Using Eqs.
(\ref{uv}) and (\ref{gap2}), after simple algebras, we rewrite the gap
(\ref{gap1}) in the following form
\begin{equation}
    \Delta_{j}=\frac{\widetilde{G}_{j}}{2}\sum_{j'}\Omega_{j'}\langle{\cal
    D}_{j'}\rangle\frac{\Delta_{j'}}{E_{j'}}~,\hspace{5mm} {\rm
    where}\hspace{5mm}
    \widetilde{G}_{j}={G}\bigg(1-G\frac{\delta{\cal
    N}_{j}^{2}}{\langle{\cal D}_{j}\rangle E_{j}}\bigg)^{-1}~.
\label{rengap}
\end{equation}
       \subsection{Finite-temperature BCS with quasiparticle number
       fluctuations}
\subsubsection{Without particle-number projection (FTBCS1)}
       The gap equation (\ref{rengap}) is remarkable as it shows that the QNF
       $\delta{\cal N}_{j}^{2}$ renormalizes the pairing interaction
       $G$ to $\widetilde{G}_{j}$.  The conventional finite-temperature
       BCS (FTBCS) gap equation $\Delta_{j}=\Delta$
       is recovered from Eq. (\ref{rengap}) when the
       following assumptions simultaneously hold:

       i) {\it Independent quasiparticles}: $n_{j}=n_{j}^{\rm FD}$, where $n_{j}^{\rm FD}$ is the Fermi-Dirac
       distribution of non-interacting fermions
              \begin{equation}
                      n_{j}^{\rm FD}=\frac{1}{e^{\beta
                      E_{j}}+1}~,
                      \label{njFD}
                      \end{equation}

       ii) {\it No quasiparticle number fluctuation}: $\delta{\cal N}^{2}_{j}=$ 0~,

       iii) {\it No screening factors}: $\langle{\cal A}_{j}^{\dagger}{\cal
              A}_{j'}^{\dagger}\rangle = \langle{\cal A}_{j}^{\dagger}{\cal
              A}_{j'}\rangle =$ 0~ in Eq. (\ref{rene}).

              These three assumptions guaranty a
              thermal quasiparticle mean field, in which
              quasiparticles are moving independently without any
              perturbation caused by the QNF and/or coupling to
              multiple quasiparticle configurations beyond the
              quasiparticle mean field. Among these
              configurations, the simplest ones are
              the small-amplitude vibrations (QRPA corrections).
From these assumptions, one can infer that releasing assumption ii) allows us
to include the effect of QNF, provided the quantal effect of coupling to QRPA
vibrations is negligible, i.e. assumption iii) still holds.
In the present article, this approximation scheme,
for which i) and iii) hold, whereas $\delta{\cal N}_{j}^{2}\neq$
0, is referred to as the FTBCS1.
\subsubsection{With Lipkin-Nogami particle-number projection (FTLN1)}
\label{FTLN1}
The problem of particle-number violation within the BCS
     theory is usually
         resolved in the simplest way by means of an approximated
         particle-number
         projection (PNP) before variation called the Lipkin-Nogami (LN)
         method~\cite{LN}.
         In Ref. \cite{SCQRPA} this method has been applied to the BCS1 and
         the resulting approach is called the LN1.  For
         the case with $\Omega_{j}\neq$ 1 and level-dependent
         gap $\Delta_{j}$ (\ref{gap1}) at $T\neq$ 0, the
         corresponding finite-temperature LN1 equations have
         the form
         \begin{equation}
        \tilde\Delta_{j}=G\sum_{j'}\Omega_{j'}\tilde\tau_{jj'}~,\hspace{3mm}
         N=2\sum_{j}\Omega_{j}\tilde\rho_{j}~,\hspace{5mm}
         \tilde\epsilon_{j}=\epsilon'_j+(4\lambda_2-G){\tilde{v}_j}^2~,\hspace{3mm}
         \lambda=\lambda_1+2\lambda_2(N+1)~,
         \label{LN1}
         \end{equation}
         where
         \begin{equation}
         \tilde{\tau}_{jj'}=\tau_{jj'}
         + \frac{2}{\Omega_{j}}\frac{\delta{\cal N}_{j}^{2}}
         {\langle{\cal D}_{j}\rangle}\delta_{jj'}\tilde{u}_{j'}\tilde{v}_{j'}~,
         \hspace{5mm} \tau_{jj'}=\langle{\cal D}_{j'}\rangle
         \tilde{u}_{j'}\tilde{v}_{j'}~,
         \hspace{5mm} \tilde{\rho}_{j}=\tilde{v}_{j}^2
         \langle{\cal D}_{j}\rangle+\frac{1}{2}(1-\langle{\cal
         D}_{j}\rangle)~,
         \label{rhotau}
         \end{equation}
         \begin{equation}
        \tilde{u}_{j}^2=\frac{1}{2}
        \left(1+\frac{\tilde{\epsilon}_{j}
        -\lambda}{\tilde{E}_{j}}\right)~,
         \hspace{5mm}
         \tilde{v}_{j}^2=\frac{1}{2}\left(1-\frac{\tilde{\epsilon}_{j}
         -\lambda}{\tilde{E}_{j}}\right)~,
         \hspace{5mm}
         \tilde{E}_{j}=\sqrt{(\tilde{\epsilon}_{j}-\lambda)^2+\tilde\Delta_{j}^2}\label{u'v'}~.
         \end{equation}
         The coefficient
         $\lambda_2$ is given as~\cite{SCQRPA}
         \begin{equation}
         \lambda_{2}=\frac{G}{4}
         \frac{\sum_{j}\Omega_{j}(1-\tilde\rho_j)\tau_{j}
         \sum_{j'}\Omega_{j'}\tilde\rho_{j'}\tau_{j'}-
         \sum_{j}\Omega_{j}(1-\tilde\rho_j)^2\tilde\rho_j^2}
         {\left[\sum_{j}\Omega_{j}\tilde\rho_j(1-\tilde\rho_j)\right]^2
         -\sum_{j}\Omega_{j}(1-\tilde\rho_j)^2\tilde\rho_j^2}~,
         \label{lambda2}
         \end{equation}
         where $\tau_{j}\equiv\tau_{jj}$.
              This FTBCS1 including the approximated PNP within the LN method
         is referred to as FTLN1 in the present article.
It is worth pointing out that, being an approximated projection that
corrects for the quantal fluctuations of particle number within the
BCS theory, the LN method in the present formulation
is not sufficient to account
for the thermal fluctuations (QNF) around the phase transition point
$T\sim T_{\rm c}$ as well as at high $T$. Another
well-known defect of the LN method is that it
     produces a large pairing gap (pairing correlation energy)
     even in closed-shell
     nuclei, where there should be no pairing gap.
     The source of this
     pathological behavior is assigned to the fast change of
     $\lambda_{2}$ at the shell closure, which invalidates the
     truncation of the expansion at second order~\cite{patho}.
In Ref. \cite{MBCS4}, it has been demonstrated
within the MBCS theory
that the projection-after-variation (PAV) method offers much better
results, which are closer to the exact solutions.
The PAV at $T\neq$ 0, however, is
much more complicated than the LN method. Therefore, we prefer to
devote a separate study to its
application to the BCS1.
\subsection{Finite-temperature BCS with quasiparticle-number
fluctuation and dynamic coupling to SCQRPA vibrations (FTBCS1+SCQRPA
and FTLN1+SCQRPA)}
As has been mentioned in the preceding section, within the quasiparticle mean field,
the expectation values $\langle{\cal A}_{j}^{\dagger}{\cal
       A}_{j'}^{\dagger}\rangle$ and $\langle{\cal A}_{j}^{\dagger}{\cal
       A}_{j'}\rangle$ at the right-hand side of Eq.
       (\ref{rene}) are always zero [Assumption iii)]. They
       cannot be factorized into the products of expectation
       values of quasiparticle-number operators within the thermal
       quasiparticle mean field
       because such crude contraction is tantamount to
       artificially breaking the pair correlators
       (\ref{A}) (See the Appendix A).
        Therefore, to account for the correlations beyond the
        quasiparticle mean field, these screening factors should be
        estimated, at least, within the SCQRPA, where
       they can be expressed below in terms of the forward- and
       backward going amplitudes, ${\cal X}^{\mu}_{j}$ and
       ${\cal Y}^{\mu}_{j}$, of the SCQRPA
       operators (phonons) as~\cite{SCQRPA}\footnote{In general,
       operator ${\cal Q}_{\mu}^{\dagger}(JM)$
       at $T\neq$ 0 also contains the terms $\sim
       B_{jj'}^{\dagger}(JM)\equiv
       [\alpha_{j}^{\dagger}\otimes\alpha_{j'}]^{J}_{M}$ and
       $B_{jj'}(JM)$ apart
       from those with ${\cal A}_{jj'}^{\dagger}(JM)$ and
       ${\cal A}_{jj'}(JM)$ because of the relation
       $\langle[B_{jj'}(JM),B_{j_{1}j_{1}'}^{\dagger}(J'M')]\rangle =
       \delta_{JJ'}\delta_{MM'}\delta_{jj_{1}}\delta_{j'j_{1}'}(n_{j}-n_{j'})\neq$
       0 for $j\neq j'$~\cite{Somer,DangJP}.
       In the present article, where $J=M=$ 0,
       and hence $j=j'$, this relation vanishes.}
       \begin{equation}
           {\cal Q}_{\mu}^{\dagger}=\sum_{j}\bigg(\frac{{\cal
           X}_{j}^{\mu}}{\sqrt{\langle{\cal D}_{j}\rangle}}{\cal
           A}_{j}^{\dagger}-\frac{{\cal
           Y}_{j}^{\mu}}{\sqrt{\langle{\cal D}_{j}\rangle}}{\cal
           A}_{j}\bigg)~,\hspace{5mm} {\cal Q}_{\mu}=[{\cal
           Q}_{\mu}^{\dagger}]^{\dagger}~.
           \label{Q}
           \end{equation}
       The renormalization factor $\sqrt{\langle{\cal
       D}_{j}\rangle}$ is introduced in Eq. (\ref{Q}) to ensure
       that the SCQRPA operators ${\cal Q}_{\mu}^{\dagger}$ and
       ${\cal Q}_{\mu}$ remain bosons within the thermal
       average (\ref{Trace}), preserving the exact commutation
       relation (\ref{[AA]}). This leads to the
       orthogonality relation for the ${\cal
       X}_{j}^{\mu}$ and ${\cal Y}_{j}^{\mu}$ amplitudes in the
       conventional form as
       \begin{equation}
           \sum_{j}\big(
           {\cal X}_{j}^{\mu}{\cal X}_{j}^{\mu'}-
           {\cal Y}_{j}^{\mu}{\cal
           Y}_{j}^{\mu'}\big)=\delta_{\mu\mu'}~,
           \label{ortho}
           \end{equation}
           which can be easily verified by calculating
           $\langle[{\cal Q}_{\mu},{\cal Q}_{\mu'}^{\dagger}]\rangle$ and
           requiring that the result to be equal to
           $\delta_{\mu\mu'}$. The inverse transformation of Eq.
           (\ref{Q}) reads
           \begin{equation}
          {\cal A}_{j}^{\dagger}=\sqrt{\langle{\cal D}_{j}}\rangle
          \sum_{\mu}\big({\cal X}_{j}^{\mu}{\cal Q}_{\mu}^{\dagger}
          +{\cal Y}_{j}^{\mu}{\cal Q}_{\mu}\big)~,
                 \label{inverse}
                 \end{equation}
provided the following conventional closure relations hold
\begin{equation}
\sum_{\mu}\big({\cal X}_{j}^{\mu}{\cal X}_{j'}^{\mu}-
{\cal Y}_{j}^{\mu}{\cal Y}_{j'}^{\mu}\big)=\delta_{jj'}~,
\hspace{5mm}
\sum_{\mu}\big({\cal X}_{j}^{\mu}{\cal Y}_{j'}^{\mu}-
{\cal Y}_{j}^{\mu}{\cal X}_{j'}^{\mu}\big)=0~,
\label{closure}
\end{equation}
\subsubsection{Screening factors}
Using the inverse transformation (\ref{inverse}), we obtain
the expectation values $\langle{\cal A}_{j}^{\dagger}{\cal
       A}_{j'}^{\dagger}\rangle$ and $\langle{\cal A}_{j}^{\dagger}{\cal
       A}_{j'}\rangle$ at $T\neq$ 0 in the form
\begin{equation}
x_{jj'}\equiv\frac{\langle{\cal A}_{j}^{\dagger}{\cal
A}_{j'}\rangle} {\sqrt{\langle{\cal D}_{j}\rangle\langle{\cal
D}_{j'}\rangle}} = \sum_{\mu}{\cal Y}_{j}^{\mu}{\cal Y}_{j'}^{\mu}
+\sum_{\mu\mu'} \bigg(U_{jj'}^{\mu\mu'}\langle{\cal
Q}^{\dagger}_{\mu} {\cal Q}_{\mu'}\rangle +
Z_{jj'}^{\mu\mu'}\langle{\cal Q}^{\dagger}_{\mu} {\cal
Q}_{\mu'}^{\dagger}\rangle\bigg)~, \label{x}
\end{equation}
\begin{equation}
y_{jj'}\equiv\frac{\langle{\cal A}_{j}^{\dagger} {\cal
A}_{j'}^{\dagger}\rangle} {\sqrt{\langle{\cal
D}_{j}\rangle\langle{\cal D}_{j'}\rangle}} = \sum_{\mu}{\cal
Y}_{j}^{\mu}{\cal X}_{j'}^{\mu} +\sum_{\mu\mu'}
\bigg(U_{jj'}^{\mu\mu'}\langle{\cal Q}^{\dagger}_{\mu} {\cal
Q}_{\mu'}^{\dagger}\rangle + Z_{jj'}^{\mu\mu'}\langle{\cal
Q}^{\dagger}_{\mu} {\cal Q}_{\mu'}\rangle\bigg)~, \label{y}
\end{equation}
where the following shorthand notations are used
\begin{equation}
    {U}_{jj'}^{\mu\mu'} = {\cal X}_{j}^{\mu}{\cal X}_{j'}^{\mu'}
       +{\cal Y}_{j'}^{\mu}{\cal
       Y}_{j}^{\mu'}~,\hspace{5mm}
    {Z}_{jj'}^{\mu\mu'} = {\cal X}_{j}^{\mu}{\cal Y}_{j'}^{\mu'}
              +{\cal Y}_{j}^{\mu'}{\cal X}_{j'}^{\mu}~.
\label{UW}
\end{equation}
taking into account the symmetry property
$\langle{\cal Q}_{\mu}^{\dagger}{\cal
Q}_{\mu'}^{\dagger}\rangle=\langle{\cal Q}_{\mu}{\cal
Q}_{\mu'}\rangle$.
Using now the definition (\ref{Q}), we express the expectation values
$\langle{\cal Q}_{\mu}^{\dagger}{\cal
Q}_{\mu'}\rangle$ and $\langle{\cal Q}_{\mu}^{\dagger}{\cal
Q}_{\mu'}^{\dagger}\rangle$
in terms of $x_{jj'}$ (i.e. $\langle{\cal
           A}^{\dagger}_{j}{\cal A}_{j'}\rangle$),
           $y_{jj'}$ (i.e. $\langle{\cal A}^{\dagger}_{j}
           {\cal A}_{j'}^{\dagger}\rangle$), and amplitudes
           ${\cal X}_{j}^{\mu}$ and ${\cal Y}_{j}^{\mu}$ as
\begin{equation}
    \langle{\cal Q}_{\mu}^{\dagger}{\cal
    Q}_{\mu'}\rangle =\sum_{j}{\cal Y}_{j}^{\mu}{\cal Y}_{j}^{\mu'}+
    \sum_{jj'}(U_{jj'}^{\mu\mu'}x_{jj'}-W_{jj'}^{\mu\mu'}y_{jj'})~,
    \label{Q+Q}
    \end{equation}
    \begin{equation}
    \langle{\cal Q}_{\mu}^{\dagger}{\cal
        Q}_{\mu'}^{\dagger}\rangle =-\sum_{j}{\cal Y}_{j}^{\mu}{\cal
        X}_{j}^{\mu'}+
        \sum_{jj'}(U_{jj'}^{\mu\mu'}y_{jj'}-W_{jj'}^{\mu\mu'}x_{jj'})~,
    \label{Q+Q+}
    \end{equation}
    where
\begin{equation}
    W^{\mu\mu'}_{jj'}={\cal X}_{j}^{\mu}{\cal Y}_{j'}^{\mu'}
              +{\cal Y}_{j'}^{\mu}
              {\cal X}_{j}^{\mu'}~.
\label{W}
\end{equation}
Inserting Eqs. (\ref{Q+Q}) and (\ref{Q+Q+})
into the right-hand sides of Eqs. (\ref{x}) and (\ref{y}),
after some simple algebras, we obtain
the following set of exact equations for the
screening factors (\ref{x}) and (\ref{y})
\[
    \sum_{j_{1}j'_{1}}\bigg[\delta_{jj_{1}}\delta_{j'j_{1}'}
    -\sum_{\mu\mu'}\big(U_{jj'}^{\mu\mu'}U_{j_{1}j_{1}'}^{\mu\mu'}
    -Z_{jj'}^{\mu\mu'}W_{j_{1}j_{1}'}^{\mu\mu'}\big)\bigg]x_{j_{1}j_{1}'}
  +\sum_{j_{1}j_{1}'\mu\mu'}\big(U_{jj'}^{\mu\mu'}W_{j_{1}j_{1}'}^{\mu\mu'}
  -Z_{jj'}^{\mu\mu'}U_{j_{1}j_{1}'}^{\mu\mu'}\big)y_{j_{1}j_{1}'}\]
  \begin{equation}
  =\sum_{\mu}{\cal Y}_{j}^{\mu}{\cal Y}_{j'}^{\mu}
  +\sum_{j''\mu\mu'}{\cal Y}_{j''}^{\mu}
  \big(U_{jj'}^{\mu\mu'}{\cal Y}_{j''}^{\mu'}
  -Z_{jj'}^{\mu\mu'}{\cal X}_{j''}^{\mu'}\big)~,
  \label{x1}
  \end{equation}
  \[\sum_{j_{1}j_{1}'\mu\mu'}\big(U_{jj'}^{\mu\mu'}W_{j_{1}j_{1}'}^{\mu\mu'}
    -Z_{jj'}^{\mu\mu'}U_{j_{1}j_{1}'}^{\mu\mu'}\big)x_{j_{1}j_{1}'}
    +\sum_{j_{1}j'_{1}}\bigg[\delta_{jj_{1}}\delta_{j'j_{1}'}
      -\sum_{\mu\mu'}\big(U_{jj'}^{\mu\mu'}U_{j_{1}j_{1}'}^{\mu\mu'}
      -Z_{jj'}^{\mu\mu'}W_{j_{1}j_{1}'}^{\mu\mu'}\big)\bigg]y_{j_{1}j_{1}'}\]
    \begin{equation}
    =\sum_{\mu}{\cal Y}_{j}^{\mu}{\cal X}_{j'}^{\mu}
    +\sum_{j''\mu\mu'}{\cal Y}_{j''}^{\mu}
    \big(Z_{jj'}^{\mu\mu'}{\cal Y}_{j''}^{\mu'}
    -U_{jj'}^{\mu\mu'}{\cal X}_{j''}^{\mu'}\big)~.
    \label{y1}
    \end{equation}

The derivation of the SCQRPA equations at finite temperature is
proceeded in the
same way as has been done at $T=$ 0,
and is formally identical to Eqs. (46), (56), and (57)
of Ref. \cite{SCQRPA} so we do not repeat them here. Notice
that the expectation values $\langle{\cal D}_{j}{\cal
D}_{j'}\rangle$ in the submatrices A and B in Eqs. (56) and
(57) of Ref. \cite{SCQRPA} are now calculated by using Eqs.
(\ref{DD}) and (\ref{QNF}). The approach that solves the
number and gap equations (\ref{N}), (\ref{gap1}) -- (\ref{QNF}),
as well as equations for the screening factors
(\ref{x1}) and (\ref{y1})
selfconsistently with the
SCQRPA ones at $T\neq$ 0, where all the assumptions i) --
      iii) cease to hold, is called the FTBCS1+SCQRPA in the
      present article. The corresponding approach that includes
      also PNP within the LN method is called as
      FTLN1+SCQRPA.
      \subsubsection{Quasiparticle occupation number}
      To complete the set of FTBCS1+SCQRPA equations we still need
      an equation for the
      quasiparticle occupation number $n_{j}$ defined in Eq.
      (\ref{nj}). Here comes the principal difference of the FTBCS1+SCQRPA
      compared to the zero-temperature SCQRPA since $n_{j}$ should be calculated
      selfconsistently from the SCQRPA taking into account
      dynamic coupling between quasiparticles and SCQRPA phonons
      at $T\neq$ 0 in an infinite hierarchy of
algebraic equations. The quasiparticle
propagator found as the formal solution of this hierarchy of equations
is different from that for free quasiparticles
by the mass operator, which reflects
the effects of coupling to complex
configurations. Since the latter cannot be treated exactly,
approximations have to be made to close the
hierarchy.  Following the same line as in Ref. \cite{DaTa},
we derive in this section a set of equations for the quasiparticle
propagator and quasiparticle occupation number $n_{j}$ at
$T\neq$ 0 by using the method of double-time Green's functions~\cite{Bogo,Zubarev}.
To close the hierarchy of equations, we lower the order of
double-time Green's functions by applying the standard
decoupling approximation introduced by Bogoliubov and Tyablikov
~\cite{Bogo,Zubarev}.

By noticing that the only term in the quasiparticle Hamiltonian
(\ref{Hqp}) that cannot be taken into account within either the BCS
theory or the SCQRPA is the sum containing $g_{j}(j')$ functionals, we
effectively rewrite Hamiltonian ${\cal H}$ in Eq. (\ref{var}) as
\begin{equation}
{\cal H}_{eff}=\sum_{j}(b'_{j}+\sum_{j'}q_{jj'}{\cal N}_{j'}){\cal N}_{j}+\sum_{\mu}\omega_{\mu}{\cal
Q}_{\mu}^{\dagger}{\cal Q}_{\mu}+\sum_{j\mu}V_{j}^{\mu}
{\cal N}_{j}({\cal Q}_{\mu}^{\dagger}+{\cal Q}_{\mu})~.
\label{Heff}
\end{equation}
The first sum at the right-hand side of this representation
describes the part of the quasiparticle Hamiltonian (\ref{Hqp}),
which cannot be expressed in terms of phonon operators (\ref{Q}).
Within the BCS theory, where the part containing $q_{jj'}$ does not
contribute whereas the term $\sim Gv_{j}^{4}$ and the QNF are neglected,
one obtains $b'_{j}=E_{j}$.
In this case, this sum corresponds to the quasiparticle mean
field.
The second sum describes the SCQRPA Hamiltonian
after solving the SCQRPA equations, which give the amplitudes
${\cal X}_{j}^{\mu}$, ${\cal Y}^{\mu}_{j}$, and the SCQRPA energies
$\omega_{\mu}$. The last sum represents the coupling between the
quasiparticle and phonon fields, which is
left out from the BCS (FTBCS1) and the QRPA (SCQRPA).
This sum is rewritten here in terms of
${\cal N}_{j}$ and  SCQRPA operators
by using the inverse transformation (\ref{inverse}). The vertex
$V_{j}^{\mu}$ obtained after this transformation has the form
\begin{equation}
V_{j}^{\mu}=\sum_{j'}g_{j}(j')\sqrt{\langle{\cal D}_{j'}\rangle}
({\cal X}_{j'}^{\mu}+{\cal Y}_{j'}^{\mu})~.
\label{V}
\end{equation}
Given that ${\cal N}_{j}$ commutes with ${\cal Q}^{\dagger}_{\mu}$ within
the SCQRPA, such effective representation of the quasiparticle
Hamiltonian causes no double counting between the first two sums at
the right-hand side of Eq. (\ref{Heff}), but becomes convenient
for the derivation of the quasiparticle Green's function, which includes the
coupling to SCQRPA modes, because the first sum
is activated only in the quasiparticle space, whereas the
second sum functions only in the phonon space.

Following closely the procedure described in Section 8.1 of Ref.
\cite{Zubarev}, we introduce the double-time retarded
Green's functions, which describe

a) {\it The quasiparticle propagation}:
\begin{equation}
G_{j}(t-t')=\langle\langle\alpha_{j}(t);\alpha^{\dagger}_{j}(t')\rangle\rangle~,
\label{Gj}
\end{equation}

b) {\it Quasiparticle-phonon coupling}:
\begin{equation}
{\Gamma}_{j\mu}^{-}(t-t')=\langle\langle\alpha_{j}(t){\cal
Q}_{\mu}(t);\alpha^{\dagger}_{j}(t')\rangle\rangle~,\hspace{5mm}
{\Gamma}_{j\mu}^{+}(t-t')=\langle\langle\alpha_{j}(t)
{\cal Q}_{\mu}^{\dagger}(t);\alpha^{\dagger}_{j}(t')\rangle\rangle~.
\label{Gamma}
\end{equation}
The magnetic quantum number $m$ in $\alpha_{jm}^{\dagger}$
and $\alpha_{jm}$ is omitted hereafter for simplicity as
the results below do not depend on $m$.
The definitions (\ref{Gj}) and (\ref{Gamma})
use the standard notation $G_{r}(t-t')=\langle\langle A(t);
B(t')\rangle\rangle\equiv-i\theta(t-t')\langle[A(t),B(t')]\rangle$ for
the double-time retarded Green's function $G_{r}(t-t')$ built from operators $A(t)$
at time $t$ and $B(t')$ at time $t'$. The advantage of using the
double-time retarded Green's function is that this type of Green's
function can be analytically continued into the complex energy plane.
The imaginary part of the mass operator in this analytic continuation
corresponds to the quasiparticle damping caused by the
quasiparticle-phonon coupling. This method is free from any
constraints of perturbation theory.

Applying the standard method of deriving the equation of motion for
the double-time Green's function, namely
\begin{equation}
    i\frac{\partial G_{r}(t-t')}{\partial t}
    =\delta(t-t')\langle[A(t),B(t)]_{\pm}\rangle +
    \langle\langle[A(t),H(t)];B(t')\rangle\rangle~,
    \label{motion}
    \end{equation}
 to the Green's functions (\ref{Gj}) and (\ref{Gamma}) with
 the effective Hamiltonian (\ref{Heff}), we find for them a set of three
 exact equations
 \begin{equation}
 i\frac{\partial G_{j}(t-t')}{\partial t} =
 \delta(t-t')+\widetilde{E_{j}}G_{j}(t-t')+\sum_{\mu}V_{j}^{\mu}\big[\Gamma_{j\mu}^{-}(t-t')+
 \Gamma_{j\mu}^{+}(t-t')\big]~,
 \label{Gjt}
 \end{equation}
 \[
  i\frac{\partial\Gamma_{j\mu}^{-}(t-t')}{\partial t} = (\widetilde{E}_{j}
  +\omega_{\mu})\Gamma_{j\mu}^{-}(t-t')+\sum_{\mu'}V_{j}^{\mu'}
  \langle\langle\alpha_{j}(t)\big[{\cal Q}_{\mu'}^{\dagger}(t)+{\cal
  Q}_{\mu'}(t)\big]{\cal Q}_{\mu}(t);\alpha^{\dagger}_{j}(t')\rangle\rangle
  \]
  \begin{equation}
  +\sum_{j'}V_{j'}^{\mu}\langle\langle\alpha_{j}(t){\cal N}_{j'}(t);
  \alpha^{\dagger}(t')\rangle\rangle~,
  \label{Gamma-}
  \end{equation}
  \[
    i\frac{\partial\Gamma_{j\mu}^{+}(t-t')}{\partial t} =
    (\widetilde{E}_{j}-\omega_{\mu})\Gamma_{j\mu}^{+}(t-t')
    +\sum_{\mu'}V_{j}^{\mu'}
    \langle\langle\alpha_{j}(t)\big[{\cal Q}_{\mu'}^{\dagger}(t)
    +{\cal
    Q}_{\mu'}(t)\big]{\cal Q}_{\mu}^{\dagger}(t);
    \alpha^{\dagger}_{j}(t')\rangle\rangle
    \]
  \begin{equation}
    -\sum_{j'}V_{j'}^{\mu}\langle\langle\alpha_{j}(t){\cal N}_{j'}(t);
    \alpha^{\dagger}(t')\rangle\rangle~,
    \label{Gamma+}
    \end{equation}
    where
    \begin{equation}
        \widetilde{E}_{j}=b'_{j}+q_{jj}~.
        \label{tildeEj}
    \end{equation}
 The last two equations, Eqs. (\ref{Gamma-}) and (\ref{Gamma+}), from
 this set contain higher-order Green's functions, which should be
 decoupled so that the set can be closed. Following
 the method proposed by Bogoliubov and
 Tyablikov~\cite{Bogo}, we decouple the higher-order Green's functions
 at the right-hand side of Eqs. (\ref{Gamma-}) and (\ref{Gamma+}) by
 pairing off operators referring to the same time, namely
 \[
     \langle\langle\alpha_{j}(t)\big[{\cal Q}_{\mu'}^{\dagger}(t)+{\cal
    Q}_{\mu'}(t)\big]{\cal Q}_{\mu}(t);
    \alpha^{\dagger}_{j}(t')\rangle\rangle \simeq
    \delta_{\mu\mu'}\nu_{\mu}G_{j}(t-t')~,\]
    \begin{equation}
    \langle\langle\alpha_{j}(t)\big[{\cal Q}_{\mu'}^{\dagger}(t)+{\cal
        Q}_{\mu'}(t)\big]{\cal Q}_{\mu}^{\dagger}(t);
        \alpha^{\dagger}_{j}(t')\rangle\rangle \simeq
        \delta_{\mu\mu'}(1+\nu_{\mu})G_{j}(t-t')~,
        \label{decouple}
        \end{equation}
        \[
\langle\langle\alpha_{j}(t){\cal N}_{j'}(t);
\alpha^{\dagger}(t')\rangle\rangle\simeq\delta_{jj'}(1-n_{j})G_{j}(t-t')~.
\]

As the result of this decoupling, Eqs. (\ref{Gamma-}) and (\ref{Gamma+}) become
 \begin{equation}
   i\frac{\partial\Gamma_{j\mu}^{-}(t-t')}{\partial t} =
   (\widetilde{E}_{j}+\omega_{\mu})\Gamma_{j\mu}^{-}(t-t')+
   V_{j}^{\mu}(1-n_{j}+\nu_{\mu})G_{j}(t-t')~,
   \label{Gammat-}
   \end{equation}
   \begin{equation}
     i\frac{\partial\Gamma_{j\mu}^{+}(t-t')}{\partial t} =
     (\widetilde{E}_{j}-\omega_{\mu})\Gamma_{j\mu}^{+}(t-t')+
     V_{j}^{\mu}(n_{j}+\nu_{\mu})G_{j}(t-t')~,
     \label{Gammat+}
     \end{equation}
     Taking the the Fourier transforms of Eqs. (\ref{Gjt}), (\ref{Gammat-}), and
     ({\ref{Gammat+}) into the (complex) energy variable $E$, one
     obtains three equations for three Green's functions $G_{j}(E)$,
     $\Gamma^{-}_{j\mu}(E)$, and $\Gamma^{+}_{j\mu}(E)$. Eliminating
     two functions $\Gamma^{\pm}_{j\mu}(E)$ by expressing them
     in terms of $G_{j}(E)$ and inserting the
results obtained into the equation for $G_{j}(E)$, we find the final equation for
the quasiparticle Green's function $G_{j}(E)$ in the form
\begin{equation}
    G_{j}(E)=\frac{1}{2\pi}\frac{1}{E-\widetilde{E}_{j}-M_{j}(E)}~,
    \label{GjE}
    \end{equation}
    where the mass operator $M_{j}(E)$ is given as
    \begin{equation}
        M_{j}(E)=\sum_{\mu}(V_{j}^{\mu})^{2}
        \bigg[\frac{1-n_{j}+\nu_{\mu}}{E-\widetilde{E}_{j}-\omega_{\mu}}+
        \frac{n_{j}+\nu_{\mu}}{E-\widetilde{E}_{j}+\omega_{\mu}}\bigg]~.
        \label{ME}
        \end{equation}
        In the complex energy plane $E=\omega\pm i\varepsilon$
        ($\omega$ real), the mass operator (\ref{ME}) can be written as
        \begin{equation}
M_{j}(\omega\pm i\varepsilon) = M_{j}(\omega)\mp
i\gamma_{j}(\omega)~, \label{M1}
\end{equation}
where
\begin{equation}
    M_{j}(\omega)=\sum_{\mu}(V_{j}^{\mu})^{2}
    \bigg[\frac{(1-n_{j}+\nu_{\mu})(\omega-\widetilde{E}_{j}-\omega_{\mu})}
    {(\omega-\widetilde{E}_{j}-\omega_{\mu})^{2}+\varepsilon^{2}}+
    \frac{(n_{j}+\nu_{\mu})(\omega-\widetilde{E}_{j}+\omega_{\mu})}
    {(\omega-\widetilde{E}_{j}+\omega_{\mu})^{2}+\varepsilon^{2}}\bigg]~,
    \label{Momega}
    \end{equation}
 \begin{equation}
     \gamma_{j}(\omega)=\varepsilon
     \sum_{\mu}(V_{j}^{\mu})^{2}
    \bigg[\frac{1-n_{j}+\nu_{\mu}}
    {(\omega-\widetilde{E}_{j}-\omega_{\mu})^{2}+\varepsilon^{2}}+
    \frac{n_{j}+\nu_{\mu}}
    {(\omega-\widetilde{E}_{j}+\omega_{\mu})^{2}+\varepsilon^{2}}\bigg]~.
    \label{gamma}
    \end{equation}

The spectral intensity $J_{j}(\omega)$ of quasiparticles is found from
the relation
\begin{equation}
  G_{j}(\omega+i\varepsilon) -
    G_{j}(\omega-i\varepsilon)=-iJ_{j}(\omega)
    (e^{\beta\omega}+1)~,
    \label{J1}
\end{equation}
and has the final form as~\cite{Bogo,Zubarev}
\begin{equation}
J_{j}(\omega)=\frac{1}{\pi}\frac{\gamma_{j}(\omega)(e^{\beta\omega}+1)^{-1}}
{[\omega-\widetilde{E}_{j}-M_{j}(\omega)]^{2}+\gamma_{j}^{2}(\omega)}~.
\label{J}
\end{equation}
Using Eq. (\ref{J}), we find the quasiparticle occupation number
$n_{j}$ as the limit $t=t'$ of the correlation function
\begin{equation}
\langle\alpha_{j}^{\dagger}(t')\alpha_{j}(t)\rangle =
\int_{-\infty}^{\infty}J_{j}(\omega)e^{-i\omega(t-t')}dt~.
\label{corr}
\end{equation}
The final result reads
\begin{equation}
    n_{j}=\frac{1}{\pi}\int_{-\infty}^{\infty}
    \frac{\gamma_{j}(\omega)(e^{\beta\omega}+1)^{-1}}
    {[\omega-\widetilde{E}_{j}-M_{j}(\omega)]^{2}+\gamma_{j}^{2}(\omega)}d\omega~.
    \label{njfinal}
    \end{equation}
In the limit of small quasiparticle damping
$\gamma_{j}(\omega)\rightarrow$ 0,
the spectral intensity $J_{j}(\omega)$ becomes a $\delta$-function, and
$n_{j}$ can be approximated with the Fermi-Dirac distribution
$[\exp(\beta {E}_{j}')+1]^{-1}$ at $\gamma_{j}({E}_{j}')\rightarrow
0$~, where $E_{j}'$ is the solution of the equation for
the pole of the quasiparticle Green's function $G_{j}(\omega)$, namely
$E'_{j}-\widetilde{E}_{j}-M_{j}(E'_{j}) = 0~$,
whereas the quasiparticle damping at $\omega=E'_{j}$ due to
quasiparticle-phonon coupling is given by
$\gamma_{j}(E'_{j})$.

We have derived a closed set of Eqs. (\ref{Momega}),
(\ref{gamma}), and (\ref{njfinal}) for the energy shift $M_{j}$,
damping $\gamma_{j}$, and occupation number $n_{j}$ of quasiparticles.
which should be solved
self-consistently with the SCQRPA equations at $T\neq$ 0 with the screening factors
calculated from Eqs. (\ref{x1}) and (\ref{y1}).
The quasiparticle occupation number $n_{j}$ obtained in this way is
used to determine the pairing gap from Eq. (\ref{gap1}).
These equations form the complete set of the FTBCS1+SCQRPA equations
for the pairing Hamiltonian (\ref{H}), where the dynamic effect of
quasiparticle-phonon coupling is self-consistently taken into account
in the calculation of quasiparticle occupation numbers.

    \section{ANALYSIS OF NUMERICAL RESULTS}
    \label{results}
    \subsection{Ingredients of calculations}

    We test the developed approach by carrying out
    numerical calculations within a schematic model
    as well as realistic single-particle spectra.
    For the schematic model, we employ the Richardson model having
    $\Omega$ doubly-folded equidistant levels with the number $\Omega$
    of levels equal to that of
    particles, $N$. This particle-hole
    symmetric case is called the half-filled one as in the absence of
    the pairing interaction ($G=$ 0), all the lowest
    $\Omega/2$ levels are occupied by $N$ particles with 2 particles
    on each level. The level distance is taken to be 1 MeV to have the
    single particle energies $\epsilon_{j}=j$ MeV with
    $j=1,\ldots,\Omega$. The results
    of calculations carried out within the FTBCS1, FTLN1, FTBCS1+SCQRPA,
    and FTLN1+SCQRPA at various $N$ and $G$ will be analyzed. As
    this model can be solved exactly~\cite{Volya}, for the sake of an illustrative
    example, we will compare the
    predictions by these approximations with the exact results
    obtained for
    $N=$ 10 and $G=$ 0.4 MeV  after extending
    the latter to finite
    temperature. Such extension is carried
    out by averaging the exact eigenvalues over
    the canonical ensemble of $N$ particles~\cite{MBCS3}.

    For the test in realistic nuclei, $^{56}$Fe and
    $^{120}$Sn, the neutron single-particle spectra for the bound states
    are obtained within the Woods-Saxon potentials
    at $T=$ 0, and kept unchanged as $T$ varies. The
    parameters of the Woods-Saxon potential for $^{120}$Sn take the following values:
    $V=$ -42.5 MeV, $V_{\rm ls}=$ 16.7 MeV, $a=a_{\rm ls}=$ 0.7 fm,
    $R=$ 6.64 fm, and $R_{\rm ls}=$ 6.46 fm. The full
    neutron spectrum for $^{120}$Sn spans an energy interval from around
    $-$37 to 7.5 MeV for $^{120}$Sn. From this spectrum
    the calculations use all 22 bound orbitals with the top bound orbital, $1i_{13/2}$,
    at energy of $-$0.478.
    For $^{56}$Fe, as we would like to compare the
    results of our approach with the predictions by the finite-temperature
    quantum Monte Carlo (FTQMC) method reported in Ref. \cite{QMC},
    the same single-particle energies from Table 1 of Ref.
    \cite{QMC} for $^{56}$Fe and the
    same values for G therein are used in calculations. Given the large number of results
    reported in Ref. \cite{QMC}, we choose to show here only one
    illustrative example for the $pf$ shell.

    The main quantities under study in the numerical analysis are
    the level-weighted gap
    \begin{equation}
        \overline{\Delta}=\frac{\sum_{j}\Omega_{j}\Delta_{j}}{\Omega_{j}}~,
      \label{avegap}
      \end{equation}
    total energy ${\cal E}=\langle{H}\rangle$, and heat capacity
    $C=\partial{\cal E}/\partial{T}$. By using PNP within the LN
    method, the internal energy has an additional term due to particle-number
    fluctuations $\Delta N^{2}$~\cite{LN}, namely
         \begin{equation}
         {\cal E}^{\rm LN1}=\langle H\rangle-\lambda_2\Delta
         N^{2}~,\hspace{5mm} \Delta N^{2}=\langle\hat{N}^{2}\rangle - N^{2}~.
         \label{ELN}
         \end{equation}
Within the FTLN1, the particle-number fluctuations $\Delta N^{2}$
consist of the quantal fluctuation, $\Delta N^{2}_{\rm QF}$, and
statistical one, $\Delta N^{2}_{\rm SF}$, which are
calculated following Eqs. (16) and (17) in Ref. \cite{DangZ},
respectively. Within the
FTLN1+SCQRPA, a term $\delta N_{\rm SC}$
due to the screening factors should be added, so that
\begin{equation}
{\Delta N^{2}}=\Delta N^{2}_{\rm QF} + \Delta N^{2}_{\rm SF}
+ \delta N_{\rm SC}~,
\hspace{5mm} \delta N_{\rm SC} =
8\sum_{jj'}\sqrt{\Omega_{j}\Omega_{j'}}
u_{j}v_{j}u_{j'}v_{j'}
             \big[\langle{\cal A}_{j}^{\dagger}{\cal A}_{j'}^{\dagger}\rangle+
             \langle{\cal A}_{j}^{\dagger}{\cal A}_{j'}\rangle\big]~.
             \label{DN2'}
             \end{equation}
    The integration in Eq. (\ref{njfinal}) is carried out within the
    energy interval $-\omega_{\rm L}\leq\omega\leq\omega_{\rm L}$
    with $\omega_{\rm L}=$ 100 MeV and a mesh point $\Delta\omega\leq$
    0.02 MeV. Since the integration limit is finite, the integral
    (\ref{njfinal}) is normalized by $\int_{-\omega_{\rm L}}
    ^{\omega_{\rm L}}J_{j}(\omega)[{\rm exp}(\beta\omega)+1]d\omega$. The results obtained within the
    FTBCS1+SCQRPA (FTLN1+SCQRPA)
    by using a smearing parameter $\varepsilon\leq$ 0.2 MeV [in calculating the
    mass operator (\ref{Momega}) and quasiparticle damping
    (\ref{gamma})] are analyzed. They remain practically the same
    with varying $\varepsilon$ up to around 0.5 MeV.
    \subsection{Results within Richardson model}
    \subsubsection{Effect of quasiparticle-number fluctuation}
    \label{Ric1}
        \begin{figure}
        \includegraphics[width=12 cm]{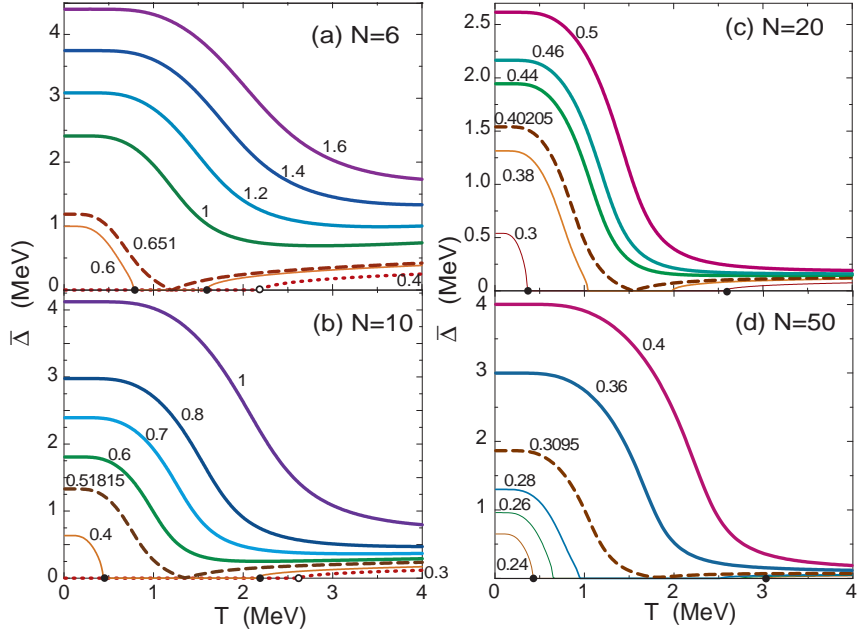}
        \caption{(Color on line) Level-weighted pairing gaps $\overline{\Delta}$
        obtained within the FTBCS1
        as functions of temperature $T$ at various values of pairing
        parameter $G$ (in MeV) indicated by the figures near the lines
        for several values of particle number $N$. Open circles on the
        axes of abscissas in panels (a) and (b) mark the values $T_{1}$
        of temperature, where
        the FTBCS1 gap turns finite at low $G$. Full circles denote
        temperature $\widetilde{T}_{\rm c}$,
        where the gap vanishes, and $T_{2}$, where it
        reappears.\label{gap}}
        \end{figure}
        It is well known that, below a critical value $G_{\rm c}$ of the pairing interaction
        parameter, the conventional BCS theory has only a trivial solution
       ($\Delta=$ 0). At $G>G_{\rm c}$, the FTBCS gap
        decreases with increasing $T$ up to a critical value of
        $T=T_{\rm c}$, where it collapses,
        and the system undergoes a sharp SN-phase transition.
        The behavior of the pairing gap within the FTBCS1 theory can be
        inferred from Eq. (\ref{rengap}).
        As a matter of fact, the
        increase of the QNF $\delta{\cal N}^{2}_{j}$ with $T$
        leads to an increase of $\widetilde{G}_{j}$, whose
        consequences are qualitatively different depending on the
        magnitude of $G$ and particle number $N$. These features can
        be seen in Fig. \ref{gap}, where
        the level-weighted pairing gaps $\overline{\Delta}$
        obtained within the FTBCS1 theory
        at various values of the pairing interaction parameter $G$
        for several particle numbers are displayed
        as functions of temperature $T$. They can be classified in
        three regions below.

        In the region of strong coupling, $G\gg G_{\rm c}$,
        where the BCS equations have non-trivial solutions at $T=0$, and
        $\delta{\cal N}^{2}_{j}$ is sufficiently large so that
        $\widetilde{G}_{j}\gg G$, the gap $\Delta_{j}$ in Eq.
        (\ref{rengap}) never collapses since whenever $T$
        reaches the value $T_{\rm c}$ where the BCS gap obtained with
        parameter $G$ collapses, the gap $\Delta_{j}$
        is always positive given
        $\widetilde{G}_{j}\gg G$ with a renormalized critical
        temperature $\widetilde{T}_{\rm c}\gg T_{\rm c}$.
        In this way, the sharp SN-phase
        transition never occurs as $\Delta_{j}$ remains always finite
        at $T_{\rm c}\leq T\ll\widetilde{T}_{\rm c}$
        with $\widetilde{T}_{\rm c}$ continuously becoming larger
        with $T$. If $G$ is sufficiently large the QNF may become so
        large at high $T$ that the level-dependent part
        $\delta\Delta_{j}$ in Eqs. (\ref{gap1}) and (\ref{gap2})
        starts to dominate and the total gap $\Delta_{j}$ will even increase
        with $T$. This effect is stronger when the particle number is
        smaller.  As seen in Fig. \ref{gap}, in contrast to the FTBCS gap, which
        collapses at $T_{\rm c}$, the FTBCS1
        gaps shown as the thick solid lines are always finite.
        For $N\geq$ 6, the gaps decrease monotonously as $T$ increases
        up to $T=$ 4 MeV.
        This feature qualitatively agrees with the
        findings within alternative approaches to thermal fluctuations
        mentioned in the Introduction.

        In the region of weak coupling,
        $G\leq G_{\rm c}$, where the pairing gap is zero at
        $T=$ 0, the increase of $\widetilde{G}_{j}$ with $T$ makes it becomes
        significantly greater than $G_{\rm c}$ at a certain $T=T_{\rm 1}$, allowing a
        non-trivial solution of the gap equation.
        This feature is demonstrated by the dotted lines in Figs.
        \ref{gap} (a) and \ref{gap} (b), where $T_{1}$ ($>$ 2 MeV)
    is marked by an open circle. Since the difference between the FTBCS1 gap $\Delta_{j}$
    and the conventional FTBCS one, $\Delta$, is the gap $\delta\Delta_{j}$ in Eqs.
    (\ref{gap1}) and (\ref{gap2}), which arises because of the QNF
    $\delta{\cal N}^{2}_{j}$, it is obvious that the finite
    gap at  $T\geq T_{1}$ is assisted by the QNF.

        In the transitional
        region, where $G$ is slightly larger than $G_{\rm c}$, it may happens
        that, although $\delta{\cal N}_{j}^{2}$ increases with $T$, it
        is still too small so that $\widetilde{G}_{j}$ is only
        slightly larger than $G$, and so is
        $\widetilde{T}_{\rm c}$ compared to $T_{\rm c}$.
        As a result, the gap collapses at
        $T=\widetilde{T}_{\rm c}$ which is slightly larger than $T_{\rm c}$.
        As $T$ increases further, the mechanism of the weak-coupling region
        is in effect, which leads to the reappearance of the gap at
        $T=T_{2}>\widetilde{T}_{\rm c}$.
        In Fig. \ref{gap}, these values
    $\widetilde{T}_{\rm c}$
    and $T_{2}$ are denoted by full circles on the axes of absiccas
    for the cases with $N=$ 6, 10, 20, 50 with $G=$ 0.6,
    0.4, 0.3, and 0.24 MeV, respectively. With increasing $G$, it
    is seen that $\widetilde{T}_{\rm c}$ increases whereas $T_{2}$ decreases so that
    at a certain $G$ these two temperatures coalesce. The
    value $G_{\rm M}$ where $\widetilde{T}_{\rm c}=T_{2} = T_{\rm M}$
    is found to be 0.651,
    0.51815, 0.40205 and 0.3095 MeV for $N=$ 6, 10, 20
    and 50, respectively, i.e. decreases with increasing $N$. The gap
    obtained with $G=G_{\rm M}$ is seen decreasing with
    increasing $T$ from 0 to $T_{\rm M}$, where it becomes zero.
    Starting from $T_{\rm M}$ the gap increases again with
    $T$. The value $T_{\rm M}$ is found increasing with $T$
    from $T_{\rm M}\simeq$ 1.2 MeV for $N=$ 6 to $T_{\rm M}\simeq$ 1.7
    MeV for $N=$ 50. At $G>G_{\rm M}$ the gap remains finite at any
    value of $T$. For small $N$, the strong
    QNF even leads to an increase of the gap with $T$ at high $T$ as
    seen in the cases with $N=$ 6, and $G_{\rm M}< G\leq$ 1.2 MeV.
    With increasing $N$ the
    high-$T$ tail of the gap gets depleted, showing how the QNF
    weakens at large $N$.

    The curious behavior of the level-weighted gap at
      weak coupling, where it appears at a certain
      $T=T_{1}$, and in the transitional region, where it collapses at
      $\widetilde{T}_{\rm c}$ and reappears at $T_{2}$, may have been
      caused by the well-known inadequacy of the BCS approximation (and BCS-based
      approaches) for weak
      pairing~\cite{Ring}. Even at $T=$ 0, Ref. \cite{Volya} has shown that,
      whereas the exact solution predicts a condensation energy of
      almost 2 MeV in the doubly-closed shell $^{48}$Ca, the BCS gives
      a normal Fermi-gas solution with zero pairing energy. It is
     expected that a proper
     PNP such as the number-projected HFB
     approach in Ref.~\cite{Sheikh}, if it can be practically extended to $T\neq$
     0, will eventually smooth out the transition points $T_{1}$ as well as
     $\tilde{T}_{\rm c}$ and $T_{2}$ in Fig. \ref{gap}.

         \begin{figure}
         \includegraphics[width=12cm]{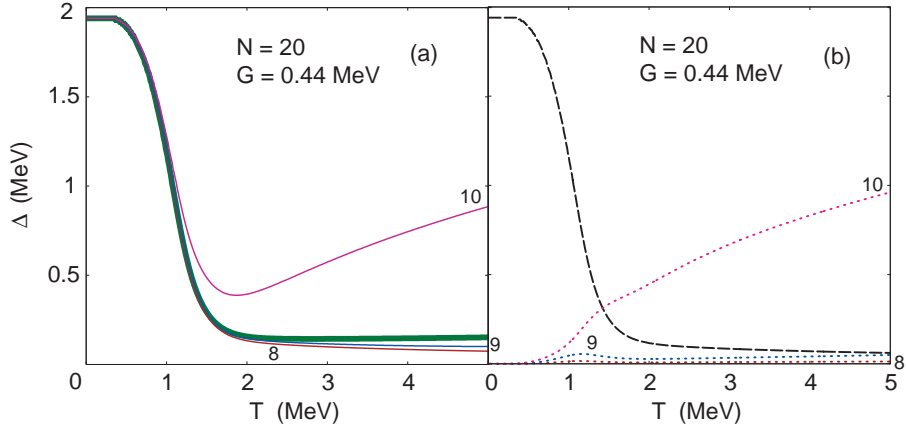}
         \caption{(Color on line)  Level-dependent pairing gap
         $\Delta_{j}$ (\ref{gap1}) and level-weighted pairing gap
         $\overline{\Delta}$ (\ref{avegap})
         obtained within the FTBCS1
         as functions of temperature $T$ for
         $N=$ 20 and $G=$ 0.44 MeV. Thick solid
         lines represent the level-weighted gaps $\overline{\Delta}$. Thin
         solid lines denote the level-dependent gaps $\Delta_{j}$
         corresponding to the $j$-th orbitals, whose level numbers $j$
         are marked at the lines. Dashed
         and dotted lines stand for the level-independent part
         (quantal component),
         $\Delta$, and
         the level-dependent one (thermal component),
         $\delta\Delta_{j}$, of the FTBCS1 gap
         $\Delta_{j}$ (\ref{gap1}), respectively. \label{gapj}}
         \end{figure}
    To have an insight into the source that causes the high-$T$ tail of the
    FTBCS1 gap we plot in Fig. \ref{gapj} the examples for the level-weighted gaps
    $\overline{\Delta}$ (\ref{avegap})
    along with the level-dependent gaps $\Delta_{j}$ (\ref{gap1}),
    which are obtained for $N=$ 20 and $G=$ 0.44 MeV.
    It is seen from this figure that
    the level-independent part (quantal component) $\Delta$ of the gap
    [dashed lines in Fig. \ref{gapj} (b)] also
    has a high-$T$ tail although it is much depleted compared to
    the total gap $\Delta_{j}$,
    which includes the level-dependent part $\delta\Delta_{j}$.
    This figure also reveals that
    the QNF has the strongest effect on the levels closest to the Fermi
    surface, which are the 10th and 11th levels. In this figure, the results
    for the 11th level are not showed as they coincide with
   those for the 10th one due to the
   particle-hole symmetry, which is well preserved within the FTBCS1.
   For the rest of levels,
   the effect of QNF is much weaker.
   With increasing the particle number $N$,
   the number of levels away from the Fermi surface becomes larger,
   whose contribution in the gap $\overline{\Delta}$
   outweighs that of the levels closest to the
   Fermi surface. This explains why
   the high-$T$ tail of the level-weighted gap $\overline{\Delta}$ is depleted
   at large $N$. When $N$ becomes very large, this tail
   practically vanishes as the total effect of QNF becomes negligible. In
   this limit, the temperature dependence of the pairing gap approaches
   that predicted by the standard BCS theory, which is well valid for
   infinite systems.
\subsubsection{Corrections due to particle-number projection and SCQRPA}
\label{schematic}
        \begin{figure}
        \includegraphics[width=12cm]{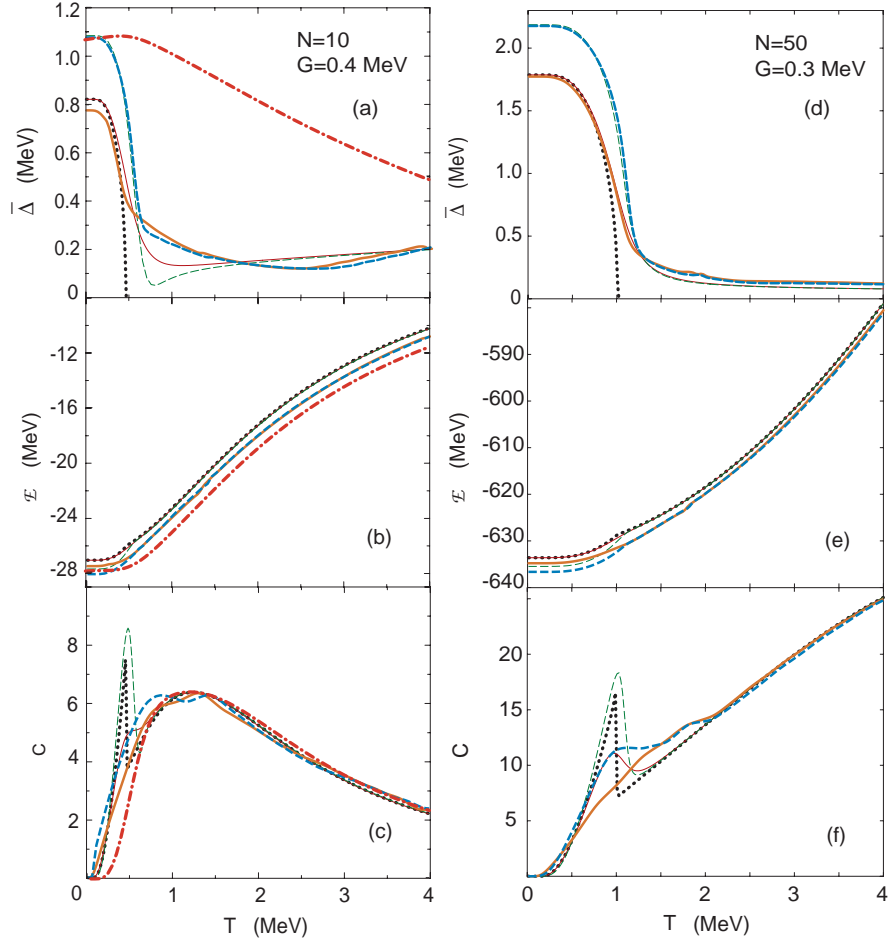}
        \caption{(Color on line)  Level-weighted pairing gaps (a, d), total
        energies (b, e),
        and heat capacities (c, f) as functions of temperature $T$, obtained
        for $N=$ 10 [(a) -- (c)], and $N=$ 50 [(d) -- (f)]. The dotted, thin solid,
        thick solid lines show the FTBCS, FTBCS1, and FTBCS1+SCQRPA
        results, respectively. The predictions by the FTLN1 and
        FTLN1+SCQRPA are presented by the thin and thick dashed
        lines, respectively. The dash-dotted lines in (a) -- (c)
        denote the exact results. The calculations of the mass operator and
        quasiparticle damping
        within the SCQRPA were performed using $\varepsilon=$ 0.05 MeV.
         \label{N10N50}}
        \end{figure}
Show in Fig. \ref{N10N50} are the level-weighted pairing gaps $\overline{\Delta}$,
total energies ${\cal E}$, and heat
capacities $C$, obtained within the FTBCS, FTBCS1, FTLN1, FTBCS1+SCQRPA, and
FTLN1+SCQRPA for the systems with $N=$ 10 ($G=$ 0.4 MeV) and
$N=$ 50 ($G=$ 0.3 MeV).
As we want to see the effect
of QNF for the case with small $\Delta(T=0)$
without any phase transition points at $\widetilde{T}_{\rm c}$ and $T_{2}$,
we choose to neglect, for this particular
test, the
self-energy term $-Gv_{j}^{2}$ in the single-particle energy.
For $N=$ 10 e.g., this increases the gap
at $T=$ 0 by around 14$\%$, to around 0.8 MeV, but the change in the
total energy is found to be negligible. Different from the common
practice, which usually neglects the
terms $\sim-G\sum_{j}\Omega_{j}v_{j}^{4}$ in calculating
the total energy ${\cal E}$, the latter
is calculated in the present article by averaging the complete pairing Hamiltonian
(\ref{Hqp}). For $N=$ 10 and $G=$ 0.4 MeV e.g.,
this causes a shift of total energy down by around 2 MeV ($\sim 8\%$)
and 1 MeV ($\sim 10.4\%$) at $T=$
0 and 4 MeV, respectively.

As has been discussed in Sec. \ref{FTLN1},
Fig. \ref{N10N50} demonstrates that, although the LN method
significantly improves the agreement between the predictions by the
FTBCS1 theory with the exact results for the pairing gap and total
energy at low $T$, it fails to do so at $T\geq T_{\rm c}$,
where all approximated results for the pairing gap
coalesce and clearly differ
from the exact result (for $N=$ 10).
The reason is partly due to the fact that, strictly speaking,
there is no pairing gap in the exact solution~\cite{MBCS3}. The dash-dotted line,
representing the exact result in Fig. \ref{N10N50}
(a) is the effective gap (canonical gap) extracted from the pairing energy.
The latter is the difference between the exact total energy and the
that of the single-particle mean field (Hartre-Fock) energies. The canonical gap
includes correlations caused by the fluctuations of the order
parameter, only a part of which is taken into account within the FTBCS1 in terms
of QNF. It reduces to
the BCS pairing gap only within the mean field approximation and the
grand canonical ensemble

The corrections caused by the SCQRPA are found to be significant
for small $N$ ($N=$ 10), in particular for the pairing gap
in the region $T_{\rm c}<T<$ 1.5 MeV [Fig. \ref{N10N50} (a)].
At $T<T_{\rm c}$, the predictions by the FTLN1+SCQRPA are closer to the exact
results than those by the FTBCS1+SCQRPA.
At $T>T_{\rm c}$ both approximations offer nearly the same results.
They produce the total
energies and heat capacities, which are much closer to the exact
values as compared to the FTBCS1 and FTLN1 results, as shown in Figs.
\ref{N10N50} (b) and \ref{N10N50} (c).
What remarkable here is that the SCQRPA correction
indeed smears out all the trace of the SN phase transition in the pairing
gap as well as energy and heat capacity.

For large $N$ ($N=$ 50), the effect of SCQRPA corrections is much smaller,
although still visible. It depletes the spike, which is the
signature of the SN
phase transition around
$T_{\rm c}$ in the heat capacity, leaving only a broad bump between
0 $<T<$ 2 MeV [Fig. \ref{N10N50} (f)].
The exact results are not available because,
for large particle
numbers, one faces technical problems of diagonalizing matrices of huge
dimension, all the eigenvalues of which should be included in the
partition function to describe correctly the total energy and heat
capacity.
        \subsection{Results by using realistic single-particle spectra}
        \label{realistic}
           \begin{figure}
           \includegraphics[width=13cm]{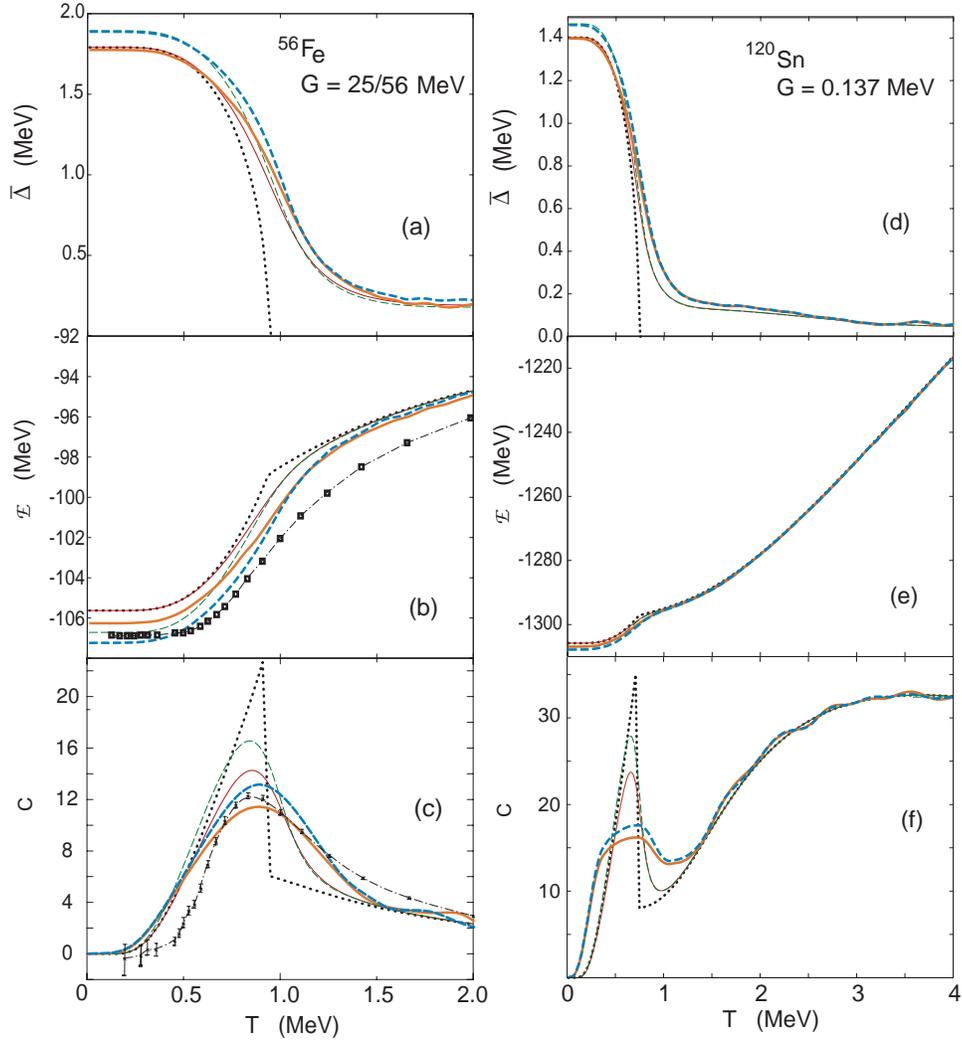}
           \caption{(Color on line)  Level-weighted pairing gaps, total
           energies, and heat capacities for
           10 neutrons in the $1f_{7/2}2p_{3/2}2p_{1/2}1f_{5/2}$ shell
           of $^{56}$Fe and all neutron bound states of
           $^{120}$Sn as functions of $T$
           ($\varepsilon=$ 0.1 MeV). Notations are
           as in Fig. \ref{N10N50}.
           In (b) and (c), the predictions by the finite-temperature
           quantum Monte Carlo method~\cite{QMC} are shown as
           boxes and crosses
           with error bars connected by dash-dotted lines.}
            \label{FeSn}
           \end{figure}
The level-weighted gaps, total energies, and heat capacities, obtained for
neutrons in $^{56}$Fe and $^{120}$Sn within the same approximations
are displayed as functions of $T$ in Fig. \ref{FeSn}. The results of calculations for 10
neutrons in the $1f_{7/2}2p_{3/2}2p_{1/2}1f_{5/2}$ shell using
$G=$ 25/26 MeV are plotted in Figs. \ref{FeSn} (a) -- \ref{FeSn}
(c) as functions of $T$ within the same temperature interval as that
in Ref. \cite{QMC}. They clearly show that the SCQRPA corrections bring the
FTBCS1 (FTLN1)+SCQRPA results closer to the predictions by the FTQMC
method for the total energy and heat capacity (No results for the pairing
gap are available within the FTQMC method in Ref. \cite{QMC}).
In heavy nuclei, such as $^{120}$Sn, the effects caused by the SCQRPA
corrections are rather small on the pairing gap and total energy.
In both nuclei, the pairing gaps do not collapse at $T=T_{\rm c}$, but
monotonously decrease with increasing $T$, and the signature of the sharp SN-phase
transition seen as a spike at $T=T_{\rm c}$ in the heat capacities is
strongly smoothed out within the FTBCS1+SCQRPA.
     \subsection{Self-consistent and statistical
     treatments of quasiparticle occupation numbers}
                \begin{figure}
                \includegraphics[width=16cm]{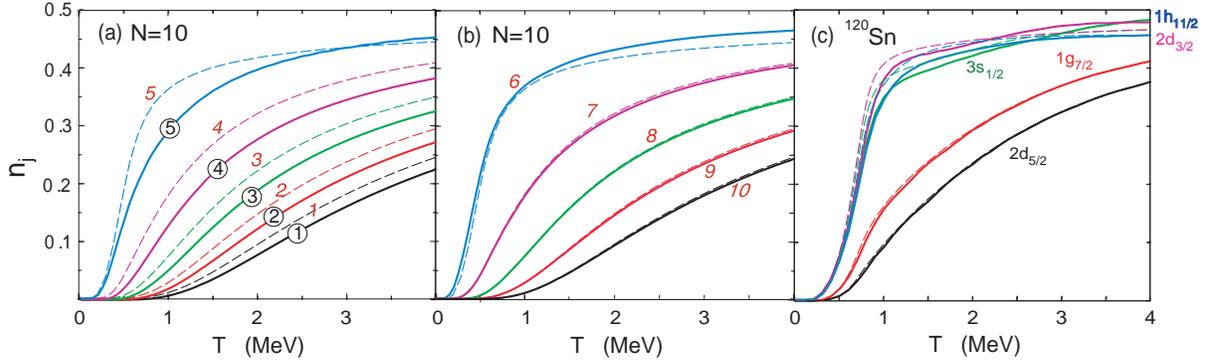}
                \caption{(Color on line) Quasiparticle
                occupation numbers
                for $N=$ 10 with $G=$ 0.4 MeV (a, b) and $^{120}$Sn
                with $G=$ 0.137 MeV (c) as functions of $T$. In (a) and (b) the
                solid lines are predictions within FTBCS1+SCQRPA
                for the levels numerated by the numbers in the circles
                starting from the lowest ones. The dashed lines,
                numerated by the italic numbers, show the
                corresponding results obtained within the
                FTBCS1. In (c) predictions for the
                neutron orbitals of the (50 - 82) shell in
                $^{120}$Sn,
                obtained within the
                FTBCS1 and FTBCS1+SCQRPA, are shown  as the
                dashed and solid lines, respectively.
            \label{njN10Sn}}
                \end{figure}
     The quasiparticle occupation numbers $n_{j}$ as predicted by the
     FTBCS1 and FTBCS1+SCQRPA for all quasiparticle levels in
     the system with $N=$ 10, $G=$
     0.4 MeV, and for the orbitals within the (50 - 82) shell in $^{120}$Sn
     ($G=$ 0.137 MeV) are shown in Fig.
     \ref{njN10Sn} as functions of $T$.
     While the $ph$ symmetry is preserved
     within the FTBCS1 ($n_{j}=n_{j}^{\rm FD}$) in the sense that
     the values for $n_{j}^{\rm FD}$ are identical for the single-particle
     levels located symmetrically from the Fermi level [Compare the dashed
     lines in Figs. \ref{njN10Sn} (a) and \ref{njN10Sn} (b)],
     it is no longer the case after taking into account dynamic
     coupling to SCQRPA vibrations. This is particularly clear in light
     systems [See the solid lines in
     Figs. \ref{njN10Sn} (a) and \ref{njN10Sn} (b)].
     This deviation of $n_{j}$ from the Fermi-Dirac distribution of free
     quasiparticles, however, turns out to be quite small
     in realistic heavy nuclei, such as $^{120}$Sn, as shown in Fig.
     \ref{njN10Sn} (c).

\subsection{Comparison between FTBCS1 and MBCS}
In Refs. ~\cite{MBCS1,MBCS2,MBCS3,MBCS4}
the MBCS theory has been developed, which
also produces a nonvanishing pairing gap at high $T$.
Therefore, it is worthwhile to draw a comparison between the MBCS theory and the
present one. Both approaches include the same QNF (\ref{QNF})
as the microscopic
source, which smoothes out the sharp SN-phase transition
and leads to the high-$T$ tail of the pairing
gap. This high-$T$ tail has been shown to be
sensitive to the size of the configuration space
in either approach.
However, due to different assumptions in these two approaches, the
functional dependences of $\delta\Delta_{j}$ on the QNF $\delta{\cal
N}_{j}^{2}$ are different. As a result,
the FTBCS1 gap is level-dependent, whereas the MBCS
one is not. The most important advantage of the FTBCS1 over the MBCS
theory is that the
solution of the FTBCS1 gap equation (\ref{gap1})
is never negative.
Moreover, at moderate and strong couplings, where the FTBCS1 gap is finite, its
behavior as a function of temperature bears no singularities in any
configuration spaces for any value of $N\geq$ 2.
The MBCS gap, on the other hand, is free from
singularities only up to a certain temperature $T_{\rm M}$, which is
around 1.75 -- 2.3 MeV within the Richardson model with $\Omega=N=$
10 and increases almost linearly with $N$ to reach
$T_{\rm M}\simeq$ 24 MeV for $\Omega=N=$ 100~\cite{MBCS3}
(For detail discussions see Refs. \cite{MBCS3,MBCS4} and references
therein).
However, the mean-field contraction
used to factorize the QNF within the FTBCS1 to the form (\ref{QNF})
may have left out some higher-order fluctuations, which can enhance
the total effect of the QNF. It might also be the reason that
causes the phase transition temperatures $T_{1}$, $\widetilde{T}_{\rm
c}$ and $T_{2}$ at weak coupling and in the transitional
region, discussed in Sec. \ref{Ric1}. Meanwhile,
the MBCS theory is based on the strict requirement of restoring the
unitarity relation for the generalized single-particle density
matrix~\cite{MHFB}, which brings in the QNF $\delta{\cal N}_{j}^{2}$
(\ref{QNF}) without the need of using a mean-field contraction.
As a result, the effect of QNF within the MBCS theory is stronger than that
predicted within the FTBCS1 and/or FTBCS1+SCQRPA, which can be
clearly seen by comparing, e.g.,
Fig.~\ref{FeSn} (d) above and Fig. 4 of Ref. \cite{MHFB}.
Whether this means that the
secondary Bogoliubov's transformation properly includes or exaggerates the
effect of coupling to configurations beyond the quasiparticle mean field
within the MBCS theory remains to be investigated. Another question
is also open on whether the MBCS theory can be improved by coupling the modified
quasiparticles to the modified QRPA vibrations. The answer to these
issues may be a subject for future study.
\section{CONCLUSIONS}
The present work extends the BCS1+SCQRPA theory, derived in Ref.
\cite{SCQRPA} for a multilevel pairing model, to finite temperature.
The resulting FTBCS1+SCQRPA theory
includes the effect of QNF as well as dynamic coupling of
quasiparticles to pairing vibrations. This theory also incorporates the
corrections caused by the particle-number projection within the LN method.

We have carried out a thorough test of
the developed approach within the Richardson model as well
as two realistic nuclei, $^{56}$Fe and $^{120}$Sn. The analysis of
the obtained pairing gaps, total energies, and heat capacities
leads to the following conclusions:

1) The FTBCS1 (with or without SCQRPA corrections) microscopically confirms
that, in the region of moderate and strong couplings,
the quasiparticle-number fluctuation smoothes out the sharp SN phase
transition, predicted by the FTBCS theory. As a result,
the gap does not collapse at $T=T_{\rm c}$, but has a
tail, which extends to high temperature $T$.

2) The correction due to the particle-number projection within
the LN method to the pairing gap is significant
at $T\ll T_{\rm c}$, which leads to a steeper temperature dependence
of the pairing gap in the region around $T_{\rm c}$.
At the same time, the SCQRPA correction smears out
the signature of a sharp SN phase transition even in heavy realistic nuclei such as
$^{120}$Sn.

3) The dynamic coupling to SCQRPA vibrations
causes the deviation of the
quasiparticle occupation number
from the Fermi-Dirac distribution for non-interacting fermions.
However, for a realistic heavy nucleus such as
$^{120}$Sn, this deviation is negligible. Consequently, in these nuclei, the FTBCS1
and FTBCS1+SCQRPA predict similar results for the pairing gap and total energy.
At the same time, for light systems,
this deviation is stronger, therefore,
the FTBCS1+SCQRPA offers a better approximation than the FTBCS1
in the study of thermal pairing properties of these nuclei.

The fact that the total energies and heat capacities obtained within
the FTBCS1+SCQRPA predictions agree reasonably well with
the exact results for $N=$ 10 as well as those
obtained within the finite-temperature
quantum Monte Carlo method for $^{56}$Fe shows that the FTBCS1+SCQRPA
can be applied in further study of thermal properties of finite
systems such as nuclei, where pairing plays an important role.
Compared to existing methods, the merit of the present approach
lies in its fully microscopic derivation and simplicity when it is
applied to heavy nuclei with strong pairing,
where the effect of coupling to SCQRPA is
negligible so that the solution of the SCQRPA can be avoided.
In this case, thermal pairing can be determined solely by solving the
FTBCS1 gap equation, which is technically as simple as the FTBCS one,
whereas the exact diagonalization is impracticable (at $T\neq$ 0).

As the next step in improving the developed approach, we will include the effect of
angular momentum in this approach. This study is now underway and the
results will be reported in a forthcoming article~\cite{Hung}.
    \begin{acknowledgments}
The authors thank Vuong Kim Au of Texas A\&M University for valuable
assistance. NQH is a RIKEN Asian Program Associate.

The numerical calculations were carried out using the {\scriptsize FORTRAN IMSL}
Library by Visual Numerics on the RIKEN Super Combined Cluster
(RSCC) system.
\end{acknowledgments}
\appendix
\label{appendix}
\section{Factorization of $\langle{\cal A}_{j}^{\dagger}{\cal
A}_{j'}\rangle$}
The factorization of the screening factor $\langle{\cal
A}^{\dagger}_{j}{\cal A}_{j'}\rangle$ is not unique as it
can be carried out in at least two ways, which lead
to different results. In the first way, one can perform
the mean-field contraction by using the Wick's
       theorem (WT) to
       obtain
       \begin{equation}
           \langle{\cal A}^{\dagger}_{j}{\cal
             A}_{j'}\rangle^{\rm WT}\simeq
             \delta_{jj'}n_{j}^{2}~.
             \label{WT}
             \end{equation}
In the second way, one uses the Holstein-Primakoff's (HP)
boson representation~\cite{HP}
\begin{equation}
    {\cal A}_{j}^{\dagger} = b_{j}^{\dagger}\sqrt{1-\frac{b_{j}^{\dagger}b_{j}}{\Omega_{j}}}~,
    \hspace{5mm}
    {\cal A}_{j} = \sqrt{1-\frac{b_{j}^{\dagger}b_{j}}{\Omega_{j}}}b_{j}~,
       \hspace{5mm} {\cal N}_{j}=2b_{j}^{\dagger}b_{j}~,
     \label{HP}
     \end{equation}
with boson operators $b_{j}^{\dagger}$ and $b_{j}$ to obtain
\begin{equation}
           \langle{\cal A}^{\dagger}_{j}{\cal
             A}_{j'}\rangle^{\rm HP}\simeq
             \delta_{jj'}\Omega_{j}n_{j}(1-2n_{j})~.
             \label{HP1}
             \end{equation}
The lowest order of the HP boson representation
implies that operators ${\cal A}_{j}^{\dagger}$
and ${\cal A}_{j}$ are ideal bosons $b_{j}^{\dagger}$
and $b_{j}$, respectively, i.e. setting ${\cal D}_{j}=$ 1 in Eq.
(\ref{[AA]}). It is in fact the well-known quasiboson approximation
(QBA), which is widely used in the derivation of the QRPA equations.
The QBA leads to
\begin{equation}
           \langle{\cal A}^{\dagger}_{j}{\cal
             A}_{j'}\rangle^{\rm QBA}\simeq
             \frac{1}{2}\delta_{jj'}\langle{\cal
             N}_{j}\rangle=\delta_{jj'}\Omega_{j}n_{j}~.
             \label{QBA}
             \end{equation}
As for the screening factor $\langle{\cal A}^{\dagger}_{j}{\cal
A}_{j'}^{\dagger}\rangle$, it vanishes in these approximations.

Using these results, we obtain the same form of Eq. (\ref{gap1}) for the
pairing gap, except that now
$\epsilon_{j}'=\epsilon_{j}$, and the
level-dependent part $\delta\Delta_{j}$ from Eq. (\ref{gap2})
       becomes
       \begin{equation}
        \delta\Delta_{j}^{\rm WT}\simeq
        2Gu_{j}v_{j}n_{j}~,
    \label{gapWT}
    \end{equation}
        \begin{equation}
        \delta\Delta_{j}^{\rm HP}\simeq
2Gu_{j}v_{j}\frac{n_{j}[1-\Omega_{j}+(2\Omega_{j}-1)n_{j}]}
{1-2n_{j}}~,
\label{gapHP}
        \end{equation}
\begin{equation}
        \delta\Delta_{j}^{\rm QBA}
        \simeq 2Gu_{j}v_{j}
        \frac{n_{j}(1-\Omega_{j}-n_{j})}{1-2n_{j}}~,
\label{gapQBA}
\end{equation}
which correspond to the approximations using the Wick's theorem, HP
representation, and the QBA, respectively.

\begin{figure}
\includegraphics[width=9cm]{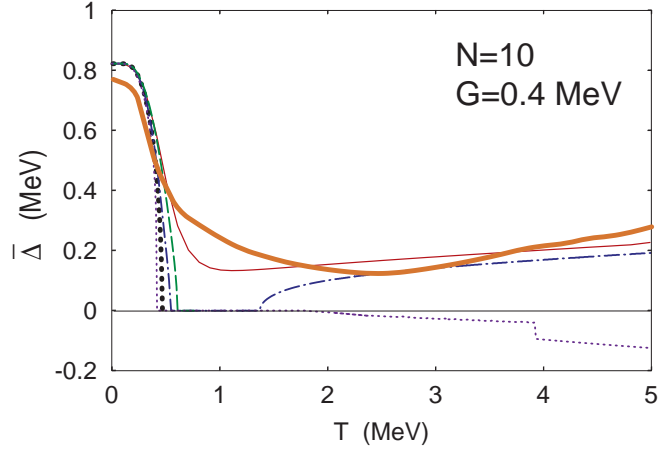}
\caption{(Color on line) Level-weighted gaps
for $N=$ 10 with $G=$ 0.4 MeV as predicted by
the WT (dashed), HP (dash-dotted), and QBA (thin dotted)
approximations in comparison with the FTBCS (thick dotted), FTBCS1
(thin solid), and FTBCS1+SCQRPA (thick solid) results.
\label{compare}}
\end{figure}
The level-weighted gaps $\overline{\Delta}$ obtained
for $N=$ 10 and $G=$ 0.4 MeV within these approximations are compared with the
FTBCS, FTBCS1 and FTBCS1+SCQRPA results in Fig. \ref{compare}.
At $T<$ 1 MeV, all three approximations, WT, HP, and QBA,
predict the gaps close to
the FTBCS one, but collapse at different $T_{\rm c}$, namely
$T_{\rm c}^{\rm QBA}<T_{\rm c}^{\rm FTBCS}<T_{\rm c}^{\rm HP}<
T_{\rm c}^{\rm WT}$. At $T\simeq$ 1.2 MeV
the HP gap reappears and increases with $T$ to reach the values
comparable with those predicted by the FTBCS1 and FTBCS1+SCQRPA at $T>$
2 MeV. From this comparison, one can see that the mean-field
contraction (\ref{WT}) for $\langle{\cal
A}_{j}^{\dagger}{\cal A}_{j'}\rangle$ includes only
a tiny fraction of the QNF because it produces a finite gap at
$T>T_{\rm c}^{\rm FTBCS}\simeq$ 0.5 MeV, but this gap
collapses again at $T_{\rm c}^{\rm WT}\simeq$ 0.6 MeV. The HP
boson representation, on the other hand, is able to take into account the
effect of QNF at hight $T$ leading to a finite gap at $T>$ 1.38 MeV,
but fails to account for this effect at intermediate temperatures
0.55 $\leq T\leq$ 1.38 MeV. The QBA
produces essentially the same result as that of the conventional
FTBCS at low $T$ with a slightly lower critical temperature
$T_{\rm c}^{\rm QBA}\simeq$ 0.43 MeV. However, it causes a negative
$\overline{\Delta}$ at $T>$ 1.9 MeV.

\end{document}